\documentclass[a4paper,11pt]{article}
\usepackage{jcappub}
\usepackage[utf8]{inputenc}
\usepackage{subfig}
\usepackage{comment}
\usepackage{appendix}
\newcommand{\phicmb}{\phi_{\text{CMB}}}
\newcommand{\ncmb}{N_{\text{CMB}}}
\newcommand{\phiend}{\phi_{\text{end}}}
\newcommand{\Trh}{T_{\text{rh}}}
\def\gsim{\:\raisebox{-0.5ex}{$\stackrel{\textstyle>}{\sim}$}\:}
\def\lsim{\:\raisebox{-0.5ex}{$\stackrel{\textstyle<}{\sim}$}\:}

\title{\Large  Large Field Polynomial Inflation:\\
	Parameter Space, Predictions and (Double) Eternal Nature}
\author[a]{Manuel Drees}
\author[a,b]{and Yong Xu}
\affiliation[a]{\it Bethe Center for Theoretical Physics and Physikalisches
	Institut, Universit\"at Bonn,\\Nussallee~12, 53115 Bonn, Germany}
\affiliation[b]{\it PRISMA+ Cluster of Excellence and Mainz Institute for Theoretical Physics, Johannes
	Gutenberg University, 55099 Mainz, Germany\footnote{Address after December 1, 2022.}}

\emailAdd{drees@th.physik.uni-bonn.de}
\emailAdd{yongxu@th.physik.uni-bonn.de}
\emailAdd{yonxu@uni-mainz.de}

\abstract{Simple monomial inflationary scenarios have been ruled out
	by recent observations. In this work we revisit the next simplest
	scenario, a single--field model where the scalar potential is a
	polynomial of degree four which features a concave ``almost'' saddle
	point. We focus on trans--Planckian field values. We reparametrize
	the potential, which greatly simplifies the procedure for finding
	acceptbale model parameters. This allows for the first comprehensive
	scan of parameter space consistent with recent Planck and BICEP/Keck
	2018 measurements. Even for trans--Planckian field values the
	tensor--to--scalar ratio $r$ can be as small as
	$\mathcal{O}(10^{-8})$, but the model can also saturate the current
	upper bound. In contrast to the small--field version of this model,
	radiative stability does not lead to strong constraints on the
	parameters of the inflaton potential. For very large field values
	the potential can be approximated by the quartic term; as well
	known, this allows eternal inflation even for field energy well
	below the reduced Planck mass $M_{\rm Pl}$, with Hubble parameter
	$H \sim 10^{-2} M_{\rm Pl}$. More interestingly, we find a region of
	parameter space that even supports {\em two phases of eternal
		inflation}. The second epoch only occurs if the slope at the
	would--be saddle point is very small, and has
	$H \sim 10^{-5} M_{\rm Pl}$; it can only be realized if
	$r \sim 10^{-2}$, within the sensitivity range of next--generation
	CMB observations.}
\begin{document} 
	\begin{flushright}
		December 2022
	\end{flushright}
	\maketitle
	\flushbottom
	
	\section{Introduction and Motivation}
	
	Inflation, as invented in the 1980s \cite{Starobinsky:1980te,
		Guth:1980zm, Linde:1981mu, Albrecht:1982wi}, is an elegant paradigm
	of early universe physics. It not only solves the horizon, flatness
	and monopole problems of standard cosmology, but also generates
	initial seeds for structure formation arising from vacuum fluctuations
	\cite{Mukhanov:1981xt}. In the simplest, ``slow--roll'' implementation
	of this idea, a spin$-0$ inflaton field is introduced which slowly
	rolls down a sufficiently flat potential; see
	Ref.~\cite{Martin:2013tda} for a comprehensive review of models of
	inflation.
	
	The simplest model assumes a monomial $\phi^p$ potential; in
	renormalizable models whose potential is bounded from below $p$ is
	either $2$ or $4$. However, such potentials are sufficiently flat,
	i.e. the first and second derivatives of the potential are
	sufficiently small compared to the potential itself, only at large
	field values. This leads to the overproduction of tensor modes,
	i.e. the tensor--to--scalar ratio $r$ is predicted beyond the upper
	bound established by recent measurements of Cosmic Microwave
	Background (CMB) anisotropies~\cite{Planck:2018vyg,
		BICEPKeck:2021gln}.\footnote{One may also consider
		monomials with fractional power, e.g. the monodromy inflationary
		model where $V(\phi) \propto \phi^{2/3}$ \cite{Silverstein:2008sg,
			McAllister:2008hb}. The BICEP/Keck 2018 results
		\cite{BICEPKeck:2021gln} then require $p < 0.53$ at $95\%$ c.l. if
		CMB scales experienced no more than 60 e--folds of inflation, in
		strong tension with monodromy inflation.}
	
	In this paper, we still assume that the inflaton is a real scalar
	field $\phi$, but we allow a general, renormalizable polynomial
	potential. This next to simplest scenario has been analyzed many
	times since 1990 \cite{Hodges:1989dw, Destri:2007pv, Aslanyan:2015hmi,
		Allahverdi:2006iq, Nakayama:2013jka, Nakayama:2013txa,
		Kallosh:2014xwa, Li:2014zfa, Gao:2015yha,
		Musoke:2017frr}. Such polynomial inflation can also
	be realized in string theory \cite{Linde:2007jn}. All these
	analyses have been performed before the release of the 2018 Planck and
	BICEP/Keck results \cite{Planck:2018vyg, BICEPKeck:2021gln}. Unlike
	previous investigations, we aim to work out the full parameter space
	that agrees with the latest measurements, and derive the allowed range
	of $r$. This is timely since future precise observations, for example
	by CORE \cite{Bouchet:2011ck}, AliCPT \cite{Li:2017drr}, LiteBIRD
	\cite{Matsumura:2013aja}, and CMB-S4 \cite{Abazajian:2019eic}, should
	greatly extend the sensitivity, down to
	$r \sim \mathcal{O} (10^{-3})$.
	
	As noted above, the potential should be flatter than a monomial at not
	too large field values. Here we achieve this by canceling several
	contributions, with different powers of the field, around a (near)
	saddle point, where both the first and second derivative of the
	potential become small. This is similar to the inflection point
	inflationary scenario \cite{Enqvist:2010vd, Hotchkiss:2011am,
		Dimopoulos:2017xox, Okada:2016ssd, Okada:2017cvy, Okada:2019yne,
		Bai:2020zil}; however, we employ a purely renormalizable potential,
	i.e. only allow terms up to $\phi^4$. Just below the would--be saddle
	point the potential has a concave shape, as favored by the Planck
	2018 data \cite{Planck:2018vyg}. We rewrite the potential in terms of
	the location $\phi_0$ of the would--be saddle point, a quantity
	$\beta$ which governs the slope of the potential at $\phi_0$, and a
	multiplicative factor which only affects the overall normalization of
	the CMB anisotropies. Based on this reparametrization, we work out,
	for the first time, the full parameter space with predictions (power
	spectrum, spectral index and its running) consistent with Planck and
	BICEP/Keck 2018 measurements \cite{Planck:2018vyg}. We find that the
	current upper bound on $r$ can be saturated, which means that part of
	the parameter space should be testable in the near future.
	
	Another aim of this work is to investigate at which scale(s) eternal
	inflation\footnote{See e.g. Ref.~\cite{Guth:2007ng} for a review.}
	might have occurred. In slow--roll inflation the classical change of
	the inflaton field during one Hubble time dominates over its quantum
	fluctuation; the inflaton field thus moves essentially
	deterministically downhill towards its minimum, until the end of
	inflation. However, in the opposite situation, where the quantum
	fluctuations dominate over the classical evolution, the inflaton field
	can move uphill rather than downhill. The Hubble patches where this
	happens inflate longer; in fact, in this case some such patches will
	inflate forever, i.e. inflation becomes eternal, although in our patch
	inflation obviously must have ended. The possibility that inflation
	can be eternal was first discussed in
	Ref.~\cite{Vilenkin:1983xq}. Later it was shown \cite{Linde:1986fc}
	that eternal inflation is in fact inevitable if the potential is a
	monomial with positive power, assuming only that the initial field
	value is sufficiently large; this occurs at energy scales well below
	the reduced Planck mass, $M_{\rm Pl} \simeq 2.4 \cdot 10^{18}$
	GeV. One thus may not need to worry about quantum gravity effects when
	describing eternal inflation \cite{Martinec:2014uva}. ``Hilltop''
	models can also lead to eternal inflation~\cite{Boubekeur:2005zm,
		Barenboim:2016mmw}.
	
	During the eternal expansion, infinitely many independent
	``mini--universes'' (or ``pocket universes'') with different de Sitter
	vacua are generated via a self-reproducing process
	\cite{Linde:1986fd}. It has been speculated that this process can
	``populate'' (or probe) the landscape of string theory
	\cite{Susskind:2003kw, Clifton:2007en, Linde:2006nw}. This mechanism
	also naturally provides a scientific justification for the (weak)
	anthropic principle. From the perspective of eternal inflation, nearly
	everything is possible, provided only that the overall energy density
	is dominated by the potential energy of the inflaton field. For
	example, independent mini--universes may feature different types of
	compactification leading to different fundamental physical laws and/or
	different values of physical ``constants'' (which are field--dependent
	in superstring theory). Some of these laws and constants support life
	of our type, and clearly we (as living beings) can only observe those
	mini--universes where this is indeed the case \cite{Linde:1986fd,
		Goncharov:1987ir}. Finally, eternal inflation may help to relax the
	initial conditions problem. By this we mean the probability that some
	initial configuration of the inflaton field, and of the other
	dynamical degrees of freedom, gives rise to sufficiently long
	exponential expansion of the universe.\footnote{For a review see
		e.g. Refs.~\cite{Goldwirth:1991rj, Brandenberger:2016uzh}.} As
	argued in \cite{Guth:2000ka}, any initial configuration that leads to
	eternal inflation will produce an infinite spacetime volume, making
	the probability of this initial configuration less significant.
	
	Since for large field values our potential is dominated by the
	$\phi^4$ term, it is not surprising that it leads to eternal inflation
	at sufficiently large values of $\phi$; the minimal required inflaton
	field energy turns out to be slightly lower than in pure $\phi^4$
	inflation.  More intriguingly, for sufficiently small values of
	$\beta$, i.e. a sufficiently flat potential near the would--be saddle
	point, a second period of eternal inflation can occur, at a value of
	the Hubble parameter smaller by more than three orders of magnitude
	than that during the first epoch of eternal inflation. However, this
	can only be realized if $\phi_0 \gsim 15 M_{\rm Pl}$, with
	$r \gsim 0.01$. These scenarios can therefore be probed in the near
	future.
	
	The remainder of this paper is organized as follows. In
	sec.~\ref{setup} we describe the general setup, with emphasis on the
	reparametrization of the inflaton potential. In
	sec.~\ref{model_parameters} methods to scan the allowed parameter
	space are shown, and the corresponding predictions are given. The
	radiative stability of the potential is checked and the maximal
	reheating temperature is determined in sec.~\ref{radiative}. In
	sec.~\ref{eternal} we investigate the possibility of realizing eternal
	inflation in our scenario, with focus on the calculation of the
	corresponding energy scale(s). Finally, sec.~\ref{sumup} summarizes
	this work. In this paper, we use Planckian units, i.e. we set the
	reduced Planck mass
	$M_{\rm Pl} = \sqrt{\frac{1}{8\pi G}} \simeq 2.4 \cdot 10^{18}$ GeV to
	unity.

	\section{The Setup} \label{setup}
	
	In this section we first describe the inflaton potential and the resulting
	expressions for the parameters of inflation; in the second subsection
	we present simplified analytical results in some limits.
	
	\subsection{General Analysis}
	
	The action for the inflaton field in the Einstein frame is given by:
	\begin{align}
	S =  \int d^4x \sqrt{-g} \left[\frac{1}{2}g_{\mu \nu}\partial^{\mu}\phi
	\partial^{\nu}\phi - V(\phi)\right]\,,
	\end{align}
	where $g$ is the determinant of the metric; we assume it to be of the
	Friedmann--Robertson--Walker (FRW) type, i.e.
	$g_{\mu \nu} = {\rm diag} (+1, -a^2, -a^2, -a^2)$ with $a$ denoting
	the scale factor. The corresponding Euler--Lagrange equation of motion
	for the classical background field is
	\begin{align}\label{eom1}
	\ddot{\phi} + 3 H \dot{\phi} -\frac{1}{a^2}  \nabla^2 \phi
	+ V^{\prime}(\phi)=0\,.
	\end{align}
	Here $\vec{\nabla}$ denotes derivatives with respect to comoving
	spatial coordinates, and $H\equiv \frac{\dot{a}}{a}$ is the Hubble
	parameter, which is determined by the Friedmann equation:
	\begin{align} \label{fri1}
	H^2 = \frac{1}{3} \rho(\phi) = \frac{1}{3} \left[\frac{1}{2} (\dot{\phi})^2
	+ V(\phi) + \frac{1}{2a^2} (\nabla \phi )^2 \right]\,.
	\end{align}
	We make the usual assumption that the classical background $\phi$ is
	homogeneous, i.e. $\phi \equiv \phi(t)$ is a function of the cosmic
	time $t$ only, so that all gradient terms for the background field
	vanish.
	The potential we are considering is the general renormalizable
	one\footnote{A linear term can be removed through a shift of $\phi$.
		We neglect the tiny cosmological constant term, which would be
		generated by a constant term in the potential.}:
	\begin{align}	\label{inflaton_potential}
	V(\phi) = b\, \phi^2 + c\, \phi^3 + d\, \phi^4 \,.
	\end{align}
	We need $d>0$ for the potential to be bound from below, and we
	consider $b> 0$ so that the minimum of the potential is at
	$\phi = \phi_{\rm min}= 0$, with $V(\phi_{\rm min}) = 0$. Since the
	potential is symmetric under the transformation $c \to -c$,
	$\phi \to -\phi$, we set $c \le 0$ without loss of generality, so that
	inflation occurs at positive field values.\footnote{The simpler case
		with $c=0$, so that the potential has only two terms, has been
		investigated in Ref.~\cite{Kobayashi:2014jga}; our analysis shows
		that this scenario is no longer viable. In
		Ref.~\cite{NeferSenoguz:2008nn}, the two-term scenario with
		radiative corrections is investigated.}  The first and second
	derivatives of the potential are given by:
	\begin{align}
	V^{\prime}(\phi) =  2 b\, \phi + 3 c\, \phi^2 + 4d\, \phi^3\,; \ \ \
	V^{\prime \prime}(\phi) =  2 b  + 6 c\, \phi + 12 d\, \phi^2\,.
	\end{align}
	We need the potential to be very flat over some range of field
	values. Suppose first that the potential features an exact saddle
	point at $\phi = \phi_0$, i.e.
	$V^{\prime}(\phi_0) = V^{\prime \prime}(\phi_0) =0$, which requires
	\begin{align} \label{correlation}
	\phi_0 = - \frac{3c}{8d}\,;  b = \frac{9c^2}{32d}\,,
	\end{align}
	from which we learn that the ratio $c/d$ determines the position of
	the saddle point. Allowing for a finite slope even at $\phi_0$, one
	can reparametrize the potential as
	\begin{align}  \label{inflaton_potential2}
	V(\phi) &=   d \left[\phi^4 +  \frac{c}{d}\left( 1- \beta  \right)\phi^3
	+  \frac{9}{32} \left(\frac{c}{d}\right)^2\phi^2\right]\nonumber \\
	& =d \left[\phi^4 +  A\left(1 - \beta \right)\phi^3
	+  \frac{9}{32} A^2\phi^2\right]\,.
	\end{align}
	Here $A\equiv \frac{c}{d} \equiv -\frac{8}{3} \phi_0$ determines the
	location of the flat region of the potential. Note that the potential
	(\ref{inflaton_potential2}) still contains three free parameters,
	$d, \, A$ and $\beta$, i.e. it is a genuine reparametrization of the
	general ansatz (\ref{inflaton_potential}). However, the form
	(\ref{inflaton_potential2}) is far more convenient, since the overall
	multiplicative factor $d$ only affects the overall normalization of
	the density perturbations, while $\beta$ directly controls the slope
	near $\phi_0$. For $\beta < 0$ the potential has a second minimum at
	$\phi > \phi_0$ where the inflaton field may get stuck, in which case
	there would be no hot Big Bang. We therefore require $\beta \geq 0$;
	recall that for $\beta = 0$ the potential has an exact saddle point at
	$\phi_0$.
	
	In this paper, we focus on the large field inflation scenario where
	$\phi_0 \geq 1$; a detailed analysis of the small field case can be found
	in Ref.~\cite{Drees:2021wgd}.
	
	The traditional potential slow--roll (SR) parameters
	\cite{Lyth:2009zz} are:
	\begin{align}\label{srparameters}
	&\epsilon_V \equiv \frac{1}{2} \left( \frac{V^{\prime}}{V}\right)^2
	=\frac {2} {\phi^2} \left[ \frac {9A^2 - 48A (\beta-1)\phi
		+ 64\phi^2} {9A^2 - 32A (\beta-1)\phi + 32\phi^2}\right]^2\,; \nonumber \\
	& \eta_V \equiv \frac{V^{\prime \prime}}{V}
	=\frac {6} {\phi^2} \left[ \frac {3A^2 - 32A (\beta-1)\phi + 64\phi^2}
	{9A^2 - 32A(\beta-1)\phi + 32\phi^2}\right]\,;  \\
	&\xi_V^2 \equiv \frac{V^{\prime} V^{\prime \prime \prime}}{V^2}
	= - \frac {384 \left[ A(\beta-1) - 4\phi\right]} {\phi^3}
	\frac {\left(9A^2 - 48A(\beta-1)\phi + 64\phi^2\right)}
	{\left(9A^2 - 32A(\beta-1)\phi + 32\phi^2\right)^2}\,.\nonumber
	\end{align}
	These quantities do not depend on $d$. During SR inflation, all these
	parameters must be small, $\epsilon_V$, $|\eta_V|$ and
	$|\xi_V^2| \ll 1$.
	
	Inflation ends at a field value $\phiend$ where
	$\epsilon_V(\phiend)=1$. Since for $\phi \ll \phi_0$ the
	term $9A^2 = 64 \phi_0^2$ dominates over other terms in
	eq.~(\ref{srparameters}), $\epsilon_V \approx 2/\phi^2 \approx \eta_V$ and
	hence $\phiend \approx 1.41$ if $\phi_0 \gg 1$. For smaller $\phi_0$,
	$\phiend$ is closer to $\phi_0$ so that the cubic and quartic terms need
	to be included in its determination.
	
	Another important quantity is the total number $\ncmb$ of e--folds of
	inflation that occurred after the CMB pivot scale
	$k_{\star} = 0.05 \rm{Mpc}^{-1}$ first crossed out the horizon; in our
	model it can be computed analytically (within the SR approximation).
	The full result is given in the Appendix; for $\beta \ll 1$ it reduces
	to
	\begin{align} \label{ncmb}
	\ncmb &= \int^{\phicmb}_{\phiend} \frac{1}{\sqrt{2 \epsilon_V}} d\phi
	\nonumber \\
	&\simeq \frac{1}{24} \left \{ 3\phi^2 - 4\phi \phi_0 + 15\phi_0^2
	-\phi_0^2 \sqrt{ \frac{2}{\beta} }
	\arctan\left( \frac{\phi_0 -\phi }{\sqrt{2\beta}\phi_0} \right)
	- \phi_0^2 \ln\left[(\phi_0 - \phi)^2 \right] \right \}
	\Bigg \vert^{\phi_{\text{CMB}}}_{\phiend}\,.
	\end{align}
	Here $\phicmb$ denotes value of the field when $k_{\star}$ crossed out
	of the horizon. In order to solve the flatness and horizon problems,
	$\gsim 50$ e--folds of inflation are needed.
	
	During SR inflation with a quasi de Sitter spacetime, Gaussian
	curvature perturbations are generated, with power \cite{Lyth:2009zz}
	\begin{equation}\label{powerspectrum}
	\mathcal{P}_{\zeta} = \frac{V}{24\pi^2\epsilon_V}\,.
	\end{equation}
	The spectral index $n_s$ and its running $\alpha$ are given by
	\begin{equation} \label{ns_and_running}
	n_s = 1 - 6\epsilon_V + 2\eta_V\,; \ \ \
	\alpha = 16\epsilon_V\eta_V - 24 \epsilon_V^2
	- 2\xi_V^2\,;
	\end{equation}
	measurements of these quantities can be used to constrain the model
	parameters $\beta$ and $A$.
	
	The final observable of interest is the tensor--to--scalar ratio $r$,
	which is given by \cite{Lyth:2009zz}
	\begin{equation} \label{ratio}
	r = 16 \epsilon_V = \frac{32}{\phi^2} \left[
	\frac {9A^2 - 48A (\beta-1)\phi + 64\phi^2}
	{9A^2 - 32A(\beta-1)\phi + 32\phi^2}\right]^2\,.
	\end{equation}
	The final Planck 2018 measurements at the pivot scale
	$k_{\star} = 0.05 \rm{Mpc}^{-1}$, including their own measurements
	plus results on baryonic acoustic oscillations (BAO), give in a 7
	parameter cosmological model (baseline $\Lambda$CDM plus running
	$n_s$) \cite{Planck:2018vyg}:
	\begin{equation}  \label{planck2018}
	\mathcal{P}_{\zeta} = (2.1 \pm 0.1) \cdot 10^{-9}\,; \
	n_s =  0.9659  \pm 0.0040\,; \  \alpha = -0.0041 \pm 0.0067\,.
	\end{equation} 
	So far no evidence for a non--vanishing tensor--to--scalar ratio $r$
	has been found. The most recent upper bound, from BICEP/Keck 2018 results
	\cite{BICEPKeck:2021gln}, is
	\begin{equation}  \label{BK2018}
	r_{0.05} < 0.035\,
	\end{equation} 
	at 95\% \text{C.L.}, after extrapolation to our pivot
	scale.\footnote{Note that in the experimental literature (e.g. the
		$r-n_s$ plots of Planck \cite{Planck:2018vyg} or BICEP/Keck
		\cite{BICEPKeck:2021gln}) the bound on $r$ is usually quoted at
		scale $k = 0.002 \rm{Mpc}^{-1} $, denoted by $r_{0.002}$. For our
		choice of the pivot scale $k_{\star} = 0.05 \rm{Mpc}^{-1}$, one has
		$r_{0.05} \simeq r_{0.002} \left(\frac{0.05}{0.002}\right)^{n_T}
		\simeq r_{0.002}\left(\frac{0.05}{0.002}\right)^{-r_{0.002}/8}$,
		where the (small) running of the tensor spectral index
		$n_T \simeq -r/8$ has been neglected.}
	
\begin{figure}[ht!]
		\centering
		\includegraphics[scale=0.8]{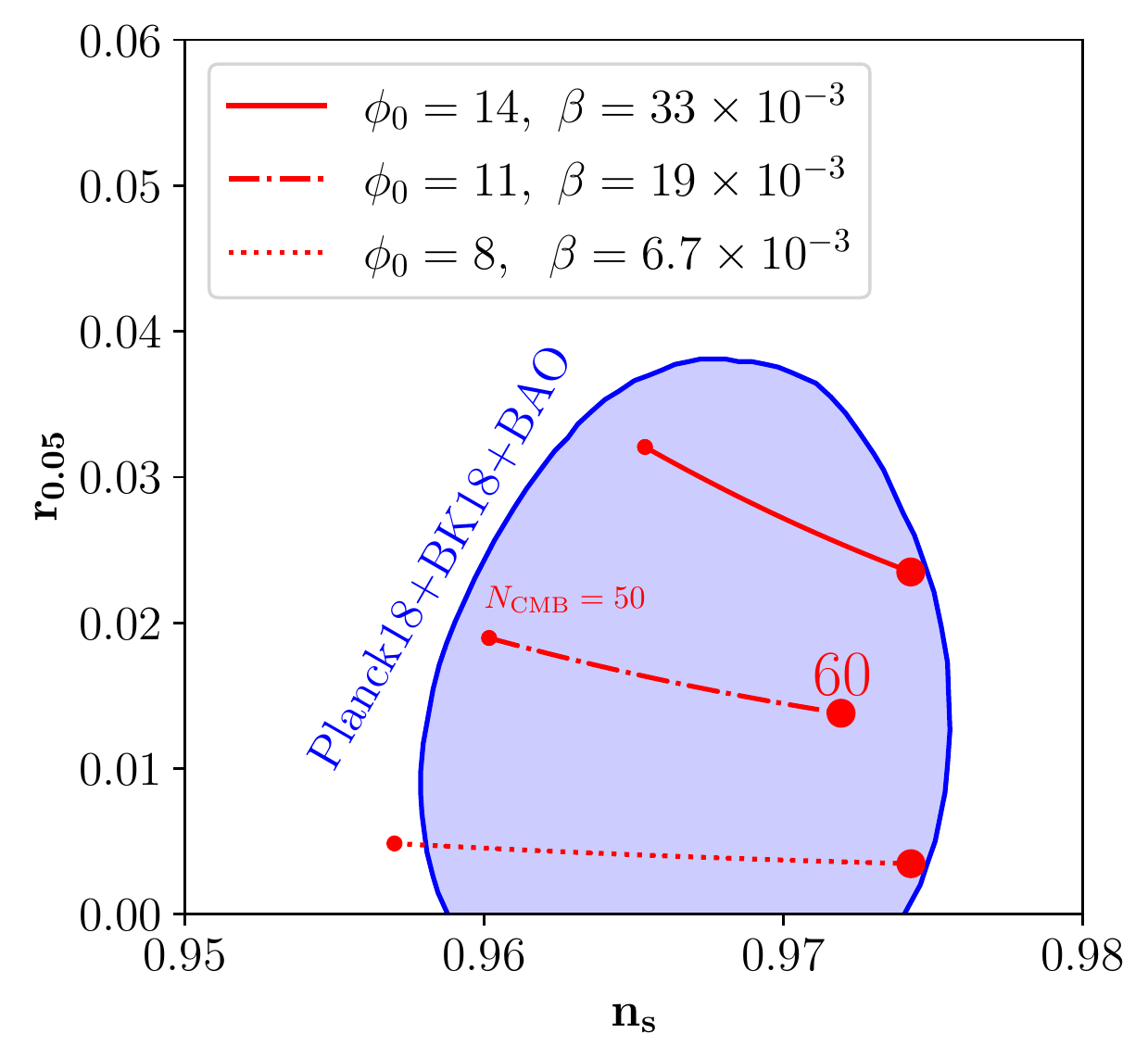}
		\caption{The blue shaded region is the currently allowed part of the
			$r-n_s$ plane; it has been adapted from the recent BICEP/Keck 2018
			results ~\cite{BICEPKeck:2021gln}. The red lines show predictions
			for our model, defined in eq.(\ref{inflaton_potential2}), where the
			parameters have been chosen such that 
			$ \mathcal{O}(10^{-3}) \lesssim r \lesssim \mathcal{O}(10^{-2})$.
			The small and big red dots correspond to $\ncmb =50 ~\text{and}~60$,
			respectively. \label{fig:r-ns}}
	\end{figure}

	The combined constraint on $r$ and $n_s$ (adapted from
	Ref.~\cite{BICEPKeck:2021gln}) is shown in Fig.~\ref{fig:r-ns}. We
	also present predictions for three sets of free parameters of the
	potential \eqref{inflaton_potential2}, chosen such that
	$\mathcal{O}(10^{-3}) \lesssim r \lesssim \mathcal{O}(10^{-2})$, which
	should be testable in the near future \cite{Bouchet:2011ck,
		Matsumura:2013aja, Li:2017drr, Abazajian:2019eic}.
	
	\subsection{Approximations}
	\label{analytical}
	
	Eqs.(\ref{srparameters}) to (\ref{ratio}) allow a fully analytical
	calculation of all SR parameters in terms of the free parameters of
	the potential plus the value of $\phicmb$, which is also a free
	parameter. However, these equations are too complicated to be solved
	analytically for the free parameters, for given $n_S$, $\alpha$ and
	$\ncmb$. In this section, we therefore present simplified analytical
	expressions, which work well in some limits.
	
	\begin{enumerate}
		
		\item $\phi \approx \phi_0$:
		
		In this regime, analytical results for the inflationary predictions
		can be obtained by rewriting the field as \cite{Drees:2021wgd}
		\begin{equation} \label{delta}
		\phi = \phi_0 (1-\delta)\,.
		\end{equation} 
		Decreasing $\phi$ corresponds to increasing $\delta$. Note that
		fluctuations at scales probed by observations of the CMB must have
		been created at $\phi < \phi_0$, where the inflaton potential is
		concave, i.e. $\eta_V < 0$, so that $n_s<1$ can be
		reproduced. Since both $\delta$ and $\beta$ are rather small (as we will
		see, $\delta \gsim {\cal O}(\sqrt{\beta})$ is needed, so that
		$\beta\ll \delta \ll 1$), it will be sufficient to keep only
		terms linear $\beta$ and up to quadratic in $\delta$ in the
		analysis.
		
		The SR parameters defined in (\ref{srparameters}) can then be
		approximated as \cite{Drees:2021wgd}:
		\begin{equation}  \label{srparameters2}
		\epsilon_V \simeq  \frac{72 \left(2\beta  + \delta^2 \right)^2}
		{\phi_0^2}; \,
		\eta_V \simeq \frac{24 \left( 2 \beta - \delta \right)} {\phi_0^2}; \,
		\xi_V^2 \simeq \frac{288 (  2\beta  + \delta^2)}{\phi_0^4} \,.
		\end{equation} 
		Using the simplified result for $\epsilon_V$, the number of
		e--folds becomes \cite{Drees:2021wgd}
		\begin{align} 	\label{ncmb2}
		N_{\rm CMB} &= \int^{\phicmb}_{\phiend} \frac{1}{\sqrt{2 \epsilon_V}} d\phi
		\nonumber \\
		& = -\frac{\phi_0^2}{12}  \int^{\delta_{\rm CMB}}_{\delta_{\rm end}}
		\frac{d \delta }{\left(2\beta  + \delta^2 \right)}\nonumber \\
		&\simeq \frac{\phi_0^2}{12\sqrt{2 \beta}}
		\left[\frac{\pi}{2} - \arctan\left(
		\frac{\delta_{\rm CMB}}{\sqrt{2 \beta}}\right) \right ]  \,.
		\end{align} 
		The normalization of the power spectrum, its spectral index and the
		running of the spectral index defined in (\ref{ns_and_running}),
		simplify to \cite{Drees:2021wgd}
		\begin{equation} \label{powerspectrum2}
		\mathcal{P}_{\zeta} 
		\simeq \frac{d \phi_0^6}{5184 \pi^2 (\delta^2 + 2\beta)^2 }\,;
		\end{equation} 
		\begin{equation}  \label{ns}
		n_s \simeq 1- \frac{ 48 \delta }{\phi_0^2}\,;
		\end{equation}
		\begin{equation}  \label{running}
		\alpha \simeq -\frac{576( 2 \beta +\delta^2)}{\phi_0^4} \,.
		\end{equation}
		Finally, the tensor--to--scalar ratio $r$ defined in eq.(\ref{ratio})
		reduces to \cite{Drees:2021wgd}
		\begin{equation} \label{ratio2}
		r \simeq \frac{1152 \left(2\beta  + \delta^2 \right)^2}{\phi_0^2} \,.
		\end{equation}

		Evidently eq.(\ref{ns}) immediately determines $\delta$,
		i.e. $\phicmb$. $\beta$ can then be fixed using eq.(\ref{ncmb2}): the
		argument of the $\arctan$ needs to be ${\cal O}(1)$, giving
		$\beta \sim {\cal O}(\delta^2)$, as claimed above. Moreover, $|\eta_V|
		\gg \epsilon_V$ in this scenario. The SR conditions thus cease to be
		satisfied at $\phiend \simeq \phi_0 \left( 1 - \phi_0^2/24 \right)$,
		where $\eta_V$ crosses $-1$ \cite{Drees:2021wgd}. However, exponential
		expansion continues until $\epsilon_V \simeq 1$, which is satisfied for
		$\phi \simeq \phi_0 \left( 1 - \sqrt{\phi_0/(6\sqrt{2} )} \right)$.
		
		In ref.\cite{Drees:2021wgd} it was shown that this approximation works
		very well for $\phi_0 \leq 1$. Here we find that it still works quite
		well even for $\phi_0 \lsim 5$.
		
		\item $\phi \gg \phi_0$:

		Here the potential is dominated by the quartic term:
		\begin{equation}
		V \simeq d \phi^4\,.
		\end{equation}
		The corresponding SR parameters are:
		\begin{equation}
		\epsilon_V =\frac{8}{\phi^2}\,;\ \ \ \eta_V = \frac{12}{\phi^2}\,,
		\end{equation}
		while the spectral index is given by
		\begin{equation}
		n_s = 1- 6\epsilon_V + 2\eta_V = 1 - \frac{24}{\phi^2}\,.
		\end{equation}
		The Planck central value $n_s=0.9659$ (cf. eq.\eqref{planck2018}) thus
		requires $\phicmb \simeq 27$, which implies
		$r=\frac{128}{\phi^2} \simeq 0.18$. This is well above the current
		upper bound on $r$, see fig.~\ref{fig:r-ns}. Moreover, $\phicmb = 27$
		with a $\phi^4$ potential yields $\ncmb \simeq 90$, which is also well
		above the allowed range. This re--derives the by now quite
		well--known result that simple $\phi^4$ inflation is excluded; as we
		noted in the Introduction, this holds for any monomial inflaton
		potential, $V(\phi) \propto \phi^p$ with $p \geq 2$.
		
		\item $\phi \gsim \phi_0$:
		
		If $\phi$ is very close to $\phi_0$ the first approximation can
		again be used, but now with $\delta < 0$. Eq.(\ref{ns}) shows that
		then $n_s > 1$ if $\beta \ll 1$, in conflict with
		observation. Solutions with $\phicmb > \phi_0$ require larger values
		of $\beta$ than those with $\phicmb < \phi_0$ in order to keep
		$\ncmb$ within the acceptable range. In fact, for $\beta > 0.1$,
		$n_s < 1$ even for $\phicmb = \phi_0$, since
		$6 \epsilon_V > 2 \eta_V$ then. However, for $\phicmb = \phi_0$ and
		$\beta = 0.1$ one has $r = 128/(9 \phi_0^2)$. $r < 0.035$ then
		requires $\phi_0 > 20$, which in turn yields\footnote{For such large
			values of $\beta$ eq.(\ref{ncmb}) is no longer valid, and the full
			expression given in the Appendix should be used.} $\ncmb >
		150$. In order to obtain a value of $n_s$ close to its upper bound
		(\ref{planck2018}) at $\phicmb = \phi_0$ one even needs
		$\beta \gsim 0.2$; $r < 0.035$ then requires $\phi_0 > 28$ and
		$\ncmb > 245$, about a factor of $4$ above the desired range.
		
		In the regime with $\phi > \phi_0$ but neither $\phi \sim \phi_0$
		nor $\phi \gg \phi_0$, the potential remains convex, i.e.
		$\eta_V > 0$. As for the case $\phicmb \simeq \phi_0$ one can still
		get $n_s < 1$, since for sufficiently large $\beta$ and/or $\phicmb$
		$\epsilon_V$ becomes larger than $\eta_V/3$. However, once again
		this leads to too large values for $\ncmb$ or $r$ (or both).
		Therefore {\it no} viable solution with $\phicmb \geq \phi_0$
		exists.
		
		\item $\phi \ll \phi_0$:
		
		If $\phi \ll \phi_0$, the quadratic term in the potential
		\eqref{inflaton_potential2} dominates:
		\begin{align}
		V(\phi) \simeq b\,\phi^2\,.
		\end{align}
		The corresponding SR parameters are
		\begin{align}
		\epsilon_V =\frac{2}{\phi^2} = \eta_V\,,
		\end{align}
		giving a spectral index
		\begin{align} \label{nsquad}
		n_s = 1- 6\epsilon_V + 2\eta_V = 1 - \frac{8}{\phi^2}\,.
		\end{align}
		The central value \eqref{planck2018} of $n_s$ measured by Planck 2018
		is reproduced for $\phicmb \simeq 15$, which in turn leads to
		$r \simeq 0.14$, well above its upper bound. ($\ncmb \simeq 56$ comes
		out correctly in this case.)
		\item $\phi < \phi_0$:
		
		In the regime with $\phi < \phi_0$ but with $\phicmb$ neither close
		to $\phi_0$ nor $\phicmb \ll \phi_0$, the potential can maintain a
		small and {\em negative} curvature due to the negative contribution
		from the cubic term slightly overcompensating the positive
		contributions from the quadratic and quartic terms. In fact, for
		$\beta \ll 1$ the potential remains concave for
		$\phi_0(1 - 2\beta) > \phi > \phi_0(1+2\beta)/3$. If $\phi_0 \lsim 1$,
		$|\eta_V|$ begins to exceed $1$, signaling the end of SR inflation,
		already at a value of $\phi$ close to $\phi_0$, see
		eq.(\ref{srparameters2}). However, for $\phi_0 > 1$ SR inflation can
		extend to field values well below $\phi_0$. In fact for very large
		$\phi_0$, the potential is effectively approaching a quadratic one
		again, since the last $\lsim 65$ e--folds of inflation happen at
		$\phi \ll \phi_0$; this leads back to the case discussed in the
		previous paragraph, which is excluded by the upper bound on
		$r$. This argument shows that there must be an upper bound on
		$\phi_0$ in our model. In the remainder of this paper we will
		explore the parameter space with $\phi_0 > 1$ and $\phicmb < \phi_0$
		in detail.
	\end{enumerate}
	
	
	\section{Model Parameters and Predictions}
	\label{model_parameters}
	
	In this section, we first describe our methods to search for
	acceptable model parameters and then scan over the full parameter
	space that is consistent with the latest CMB observations
	(\ref{planck2018}) and \eqref{BK2018} at the $2\sigma$ level.
	
	\subsection{Method to Find Model Parameters and Examples}
	\label{method}
	
	We have learned that the location of the plateau, i.e. $\phi_0$, is
	determined by the parameter $A$ of the rewritten potential
	(\ref{inflaton_potential2}); we treat it as a free parameter. The
	slope of the plateau is determined by $\beta$, i.e. for given field
	value (not too far from $\phi_0$) the SR parameter $\epsilon_V$ will
	become larger when $\beta$ is increased. Of course, the SR parameters,
	as well as $\ncmb$ and $r$, also depend on $\phicmb$. On the other
	hand, the overall coupling $d$ in eq.(\ref{inflaton_potential2}) only
	affects the normalization of the power spectrum, see
	eq.(\ref{powerspectrum}).
	
	In practice we first fix $\phi_0$. The parameters $\phicmb$ and
	$\beta$ should then be chosen such that $n_s$ and $\ncmb$ have the
	desired values. As argued in the previous subsection, viable solutions
	only exist for $\phicmb < \phi_0$. Reducing $\phicmb$ for given
	$\beta$ means that one is moving away from the flattest part of the
	potential (apart from the region near the minimum, which cannot lead
	to inflation); this increases $\epsilon_V$ and reduces $\eta_V$ (often
	making it more negative). This means that $1-n_s$ and $r$ both
	increase when $\phicmb$ is reduced, but $\ncmb$ becomes
	smaller. Reducing $\beta$ for fixed $\phicmb$ has the opposite effect:
	the potential becomes flatter, which increases $\ncmb$ but decreases
	$r$ and usually also $1-n_s$.
	
	For $\phi_0 \lsim 5$ the approximation described by
	eqs.(\ref{delta})--(\ref{ratio2}) still works fairly well. As already
	noted in the corresponding discussion, in this case one can use the
	spectral slope $n_s$ to determine $\phicmb$, and then chose $\beta$
	such that $\ncmb$ is reproduced. For this range of parameters $r$ is
	still very small, well below the present bound.
	
	This procedure yields $1 - \phicmb/\phi_0 \propto \phi_0^2$, so the
	approximation $\phi_0 - \phicmb \ll \phi_0$ begins to break down for
	$\phi_0 \geq 5$. The free parameters $\phicmb$ and $\beta$ then need
	to be determined together. We find that this can still be done
	iteratively.  One starts with a guess for $\beta$, e.g. the
	small$-\phi_0$ value $\simeq 10^{-6} \phi_0^4$. For this value of
	$\beta$, $\phicmb$ is selected such that the spectral index $n_s$
	comes out as desired. One then fixes $\phicmb$ and varies $\beta$
	until $\ncmb$ takes the desired value. With this new value of $\beta$,
	a new value of $\phicmb$ can be computed using $n_s$, and so on. This
	iteration usually converges fairly quickly. At the end, the overall
	coupling strength $d$ is determined using eq.(\ref{powerspectrum})
	with ${\cal P}_\zeta = 2.1 \cdot 10^{-9}$.
	
	Of course, one should also check that $r$ and the
	running of the spectral index $\alpha$ have acceptable values. We find
	that $\alpha$ is always negative, and lies within the currently allowed
	range given in (\ref{planck2018}). On the other hand, for large $\phi_0$
	the tensor--to--scalar ratio $r$ may come out too large. Moreover, while
	for sufficiently small $\phi_0$ the desired values of $n_s$ and $\ncmb$
	can always be attained, this is not necessarily true for larger values
	of $\phi_0$.
	
	\begin{figure}[ht!]
		\centering
		\includegraphics[scale=0.6]{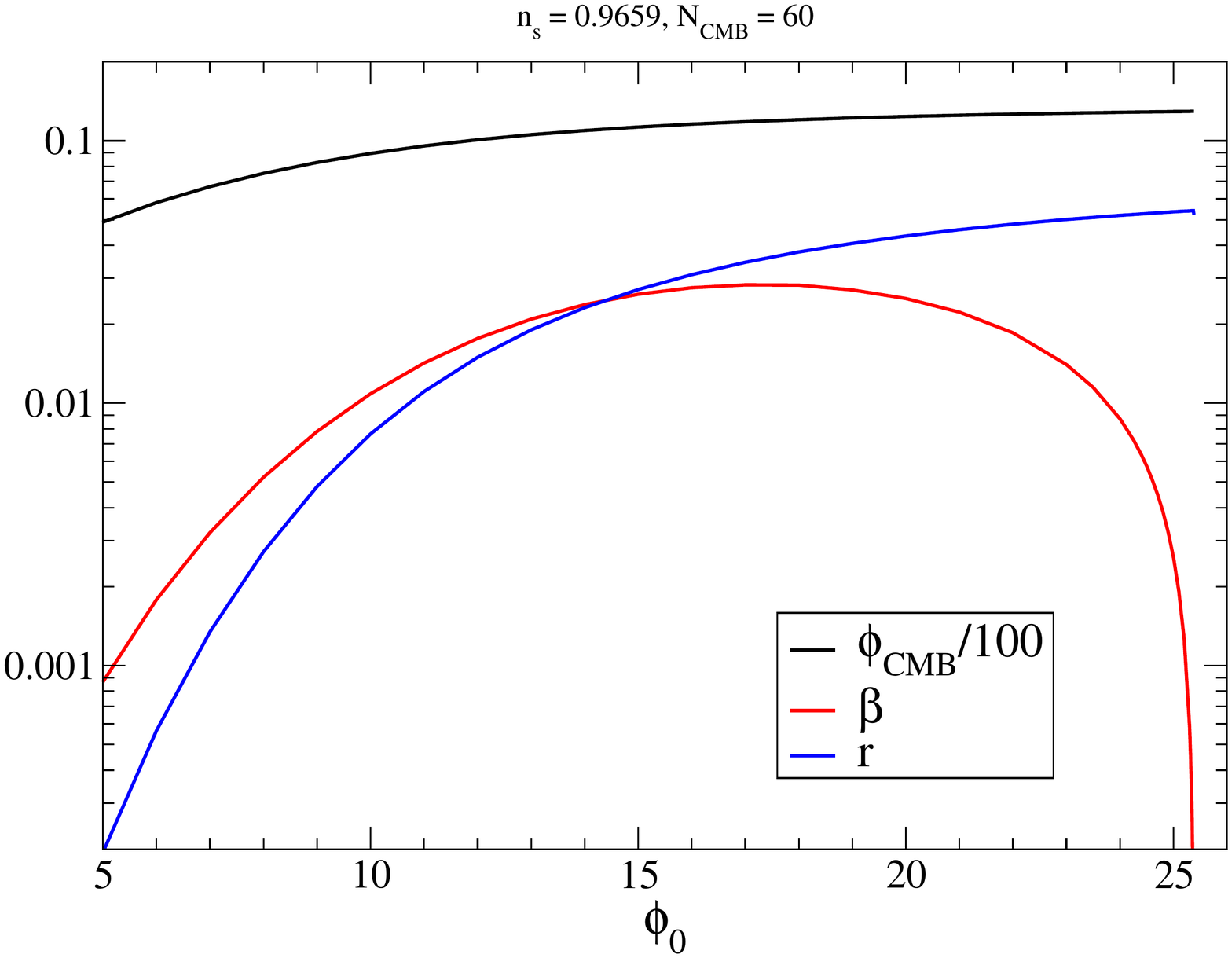}
		\caption{The values of $\phicmb$ (black, divided by $100$) and
			$\beta$ (red) that lead to $n_s = 0.9659$ and $\ncmb = 60$, as
			function of $\phi_0$. The blue curve shows the resulting
			prediction for the tensor--to--scalar ratio $r$.\label{fig:sample}}
	\end{figure}
	
	This procedure is illustrated in Fig.~\ref{fig:sample}, for
	$\phi_0 \geq 5$ where the deviation from the small$-\phi_0$ solution
	begins to be sizable. Here we have chosen $\ncmb = 60$ and
	$n_s = 0.9659$, the current central value. We see that for
	$\phi_0 \lsim 10$, $\phicmb$ (shown in black) has to remain quite
	close to $\phi_0$. For fixed ratio $\phicmb/\phi_0$ an increase of
	$\phi_0$ reduces the SR parameters due to the overall $1/\phi^2$
	factors in eqs.(\ref{srparameters}), moving $n_s$ closer to 1. This
	has to be compensated by decreasing $\phicmb/\phi_0$. On the other
	hand, $\ncmb$ depends not only on the SR parameter $\epsilon_V$, but
	also on the range of field values over which the integral in
	eq.(\ref{ncmb}) has to be evaluated. An increase of $\phi_0$ thus has
	to be compensated by an increase in $\beta$, shown by the red curve,
	in order to leave $\ncmb$ unchanged. This leads to a rapid increase
	of $\epsilon_V$, and hence of $r$ (shown in blue); however, in this
	region of parameter space we still have $\epsilon_V \ll |\eta_V|$,
	i.e. the spectral index $n_s$ is essentially determined by $\eta_V$.
	
	For $\phi_0 > 10$ the curve for $\phicmb$ flattens out. Recall that
	even for a purely quadratic potential $\phicmb \simeq 15$, see
	eq.(\ref{nsquad}), and our potential is significantly flatter, hence
	requiring smaller $\phicmb$ in order to give the correct $\ncmb$.
	At the same time $\epsilon_V$ keeps increasing, so that its contribution
	to $n_s$ becomes significant. This flattens the increase of $\beta$,
	which reaches a maximum at $\phi_0 \simeq 17$. For yet larger values
	of $\phi_0$, $\phicmb$ becomes almost independent of $\phi_0$, i.e.
	inflation now occurs further and further away from the saddle point.
	Keeping the potential sufficiently flat then requires a {\em reduction}
	of $\beta$. For the given choice of $n_s$ and $\ncmb$ no solution can
	be found for $\phi_0 \geq 25.4$; the red curve drops very steeply at
	the end, since the potential at $\phicmb \sim 0.55 \phi_0$ depends only weakly
	on the slope at $\phi_0$.
	
	Note that $r$ keeps increasing even when $\beta$ is decreasing; for
	the chosen parameters, it exceeds the bound of $0.035$ for $\phi_0 \geq
	17.5$. However, for other choices of $\ncmb \in [50,65]$ and
	$n_s \in [0.9579,0.9739]$ (the $2\sigma$ range) the solution terminates
	before the bound on $r$ is saturated.\footnote{As long as only field
		values below $\phi_0$ are considered, one could find additional solutions
		with larger $\phi_0$ and negative $\beta$. Recall, however, that in this
		case the potential features a second minimum above $\phi_0$, in which
		the inflaton would get stuck if it came from yet larger field values;
		in particular, in scenarios featuring a phase of eternal inflation, as
		discussed below. This is why we restrict ourselves to $\beta \geq 0$.}
	Notice also that $\beta$ remains quite small throughout, i.e. the potential
	indeed needs to feature a near--inflection point.
	
	\begin{table}[!t]
		\centering
		\begin{tabular}{|| c c  c c||  c c c c||} 
			\hline
			$\phi_0$ & $d/10^{-14} $  & $\beta / 10^{-3}$& $\phicmb$ & $n_s$ & $r /10^{-3}$
			& $\alpha / 10^{-3}$& $N_{\rm{CMB}}$
			\\ [0.5ex] \hline\hline
			$1$ & $0.00636$  & $0.0009$ & $0.999203$&  $0.9619$ & $6.81\cdot 10^{-6}$
			& $-1.40$& $64.6$\\
			$1$ & $0.164$  & $0.0017$ & $0.999285$&  $0.9659$ & $1.76 \cdot 10^{-5}$
			& $-2.24$  & $54.6$\\
			$1$ & $0.268$  & $0.0023$ & $0.999368$&  $0.9699$ & $2.87 \cdot 10^{-5}$
			& $-2.87$& $50.2$
			\\ \hline
			$2$ & $0.27$  & $0.015$ & $1.99356$&  $0.9619$ & $4.64\cdot10^{-4}$ & $-1.43$
			& $64.3$\\
			$2$ & $0.65$  & $0.027$ & $1.99418$&  $0.9659$ & $1.11 \cdot 10^{-3}$
			& $-2.22$& $55.2$\\
			$2$ & $1.03$  & $0.036$ & $1.99482$&  $0.9699$ & $1.77\cdot10^{-3}$ & $-2.80$
			& $51.1$
			\\ \hline
			$3$ & $0.66$  & $0.08$ & $2.97783$&  $0.9619$ & $5.74 \cdot 10^{-3}$
			& $-1.47$& $63.8$\\
			$3$ & $1.52$ & $0.14$ & $2.97976$ & $0.9659$ & $1.32 \cdot 10^{-2}$ & $-2.24$
			& $55.2$\\
			$3$ & $2.49$ & $0.19$ & $2.98175$ & $0.9699$ & $2.17 \cdot 10^{-2}$ & $-2.87$
			& $50.7$
			\\ \hline
			$4$ & $1.16$ & $0.25$ & $3.9461$ & $0.9619$ & $3.18\cdot 10^{-2}$ & $-1.43$
			& $64.4$\\
			$4$ & $2.77$ & $0.45$ & $3.9499$ & $0.9659$ & $7.62\cdot 10^{-2}$ & $-2.22$
			& $55.2$\\
			$4$ & $4.39$ & $0.60$ & $3.9542$ & $0.9699$ & $1.21 \cdot 10^{-1}$ & $-2.81$
			& $51.0$
			\\ \hline
			$5$ & $2.23$  & $0.7$ & $4.8899$& $0.9619$ & $0.15$ & $-1.55$& $61.9$\\
			$5$ & $4.29$ & $1.1$ & $4.8968$& $0.9659$ & $0.29$ & $-2.16$& $55.2$\\
			$5$ & $7.02$ & $1.5$ & $4.9037$&  $0.9699$ & $0.48$ & $-2.77$& $50.6$
			\\ \hline
			$8$ & $4.54$ & $4.0$ & $7.4815$ & $0.9619$ & $2.04$ & $-1.23$& $62.3$\\
			$8$ & $8.76 $  & $6.7$ & $7.4879$ & $0.9659$ & $4.01$ & $-1.72$& $55.0$\\
			$8$ & $13.05$  & $9.0$ & $7.5048$ & $0.9699$ & $6.07$ & $-2.12$& $51.2$
			\\ \hline
			$11$ & $5.19$ &$10$ & $9.4906$ & $0.9619$ & $8.5$ & $-0.84$& $62.1$\\ 
			$11$ & $9.43$  & $19$ & $9.4772$ & $0.9659$ & $16.1$ & $-1.18$& $54.7$\\
			$11$ & $11.99$  & $25$ & $9.5432$ & $0.9699$ & $21.1$ & $-1.35$& $52.7$
			\\ \hline
			$14$ & $6.93$ &$26$ & $10.6090$ & $0.9619$ & $29.8$ & $-0.82$& $53.7$\\
			$14$ & $7.05$  & $33$ & $10.8235$ & $0.9659$ & $31.5$ & $-0.83$& $55.3$\\
			$14$ & $7.36$  & $41$ & $11.0439$ & $0.9699$ & $34.2$ & $-0.86$& $56.6$
			\\ \hline
			$17$ & $3.28$ &$8$ & $11.6109$ & $0.9619$ & $27.1$ & $-0.49$& $61.1$\\
			$17$ & $3.43$  & $22$ & $11.8913$ & $0.9659$ & $30.0$ & $-0.51$& $62.6$\\
			$17$ & $3.73$  & $39$ & $12.1859$ & $0.9699$ & $34.9$ & $-0.55$& $63.6$
			\\ \hline
			$20$ & $2.25$ &$3$ & $12.4500$ & $0.9651$ & $32.9$ & $-0.42$& $64.3$\\
			$20$ & $2.28$  & $8$ & $12.5183$ & $0.9659$ & $33.9$ & $-0.42$& $64.6$\\
			$20$ & $2.30$  & $13$ & $12.6000$ & $0.9668$ & $34.7$ & $-0.43$& $65.0$
			\\ \hline \hline
		\end{tabular}
		\caption{Examples of model parameters and corresponding predictions.
			The overall coupling strength $d$ has been chosen to reproduce the
			central value of power spectrum, i.e.
			$\mathcal{P}_{\zeta} \simeq 2.1 \cdot 10^{-9}$. The predictions for
			$n_s$ and $\alpha$ are consistent with Planck 2018 results
			(\ref{planck2018}) at the $1\sigma$ level. Predictions for the
			tensor--to--scalar ratio $r$ satisfy the current bound $r< 0.035$
			(from the recent BICEP/Keck 2018 \cite{BICEPKeck:2021gln}) and range
			from $\mathcal{O}(10^{-8})$ to $\mathcal{O}(10^{-2})$.}
		\label{table:1}
	\end{table}
	
	In Table~\ref{table:1} we explore a wider range\footnote{Generically
		$N_{\rm CMB}<65$ unless there is an exotic reheating phase following
		the end of inflation \cite{Liddle:2003as,Tanin:2020qjw}. For our
		case, inflation ends with a usual quadratic potential, hence we adopt
		$50 \leq \ncmb \leq 65$.} of $\ncmb$ and $n_s$, for $\phi_0$
	between $1$ and $20$. The overall trends are as in
	Fig.~\ref{fig:sample}: $\phicmb$ remains close to $\phi_0$ for
	$\phi_0 \lsim 10$, but increases only slowly once $\phi_0 > 15$; and
	$\beta$ at first increases quickly, but reaches a maximum at $\phi_0$
	around $15$ and then quickly diminishes again. We also see that the
	model can saturate the upper bound on $r$ for $\phi_0 \geq 14$.
	
	Moreover, the running $\alpha$ of the spectral index is always
	negative, and well within the currently allowed range given in
	(\ref{planck2018}). It is essentially independent of $\phi_0$ for
	$\phi_0 \lsim 5$, but becomes smaller in magnitude for
	$\phi_0 \geq 8$. In fact, $\xi_V^2$ changes sign at
	$\phicmb = 2 \phi_0 (1- \beta)/3$; for the parameters used in
	Fig.~\ref{fig:sample} this happens at $\phi_0 \simeq 18$. However, by
	then $\epsilon_V$ has become so large that the first two terms in the
	expression for $\alpha$ in eq.(\ref{ns_and_running}) dominate; these
	terms are negative since $\eta_V < 0$.
	
	Table~\ref{table:1} also lists the values of $d$ needed to reproduce
	the observed normalization of the spectrum of CMB anisotropies,
	computed from eq.(\ref{powerspectrum}). At small $\phi_0$ the coupling
	scales like $d \propto \phi_0^2$, since
	$V(\phicmb) \propto d \phi_0^4$ and $\epsilon_V \propto
	\phi_0^6$. However, for $\phi_0 > 5$ the growth of $\epsilon_V$ with
	increasing $\phi_0$ slows down. As a result, $d$ reaches a maximum
	value near $10^{-13}$ at $\phi_0 \sim 10$, and decreases again for yet
	larger values of $\phi_0$. As a result, the physical inflaton mass
	$m_\phi = 2 \sqrt{d} \phi_0$ increases $\propto \phi_0^2$ for
	$\phi_0 \lsim 5$, but depends only weakly on $\phi_0$ for
	$\phi_0 > 10$.

	\subsection{Complete Scan of Parameter Space}
	\label{sec_ps}
	\begin{figure}[ht!]
		\centering
		\includegraphics[scale=0.8]{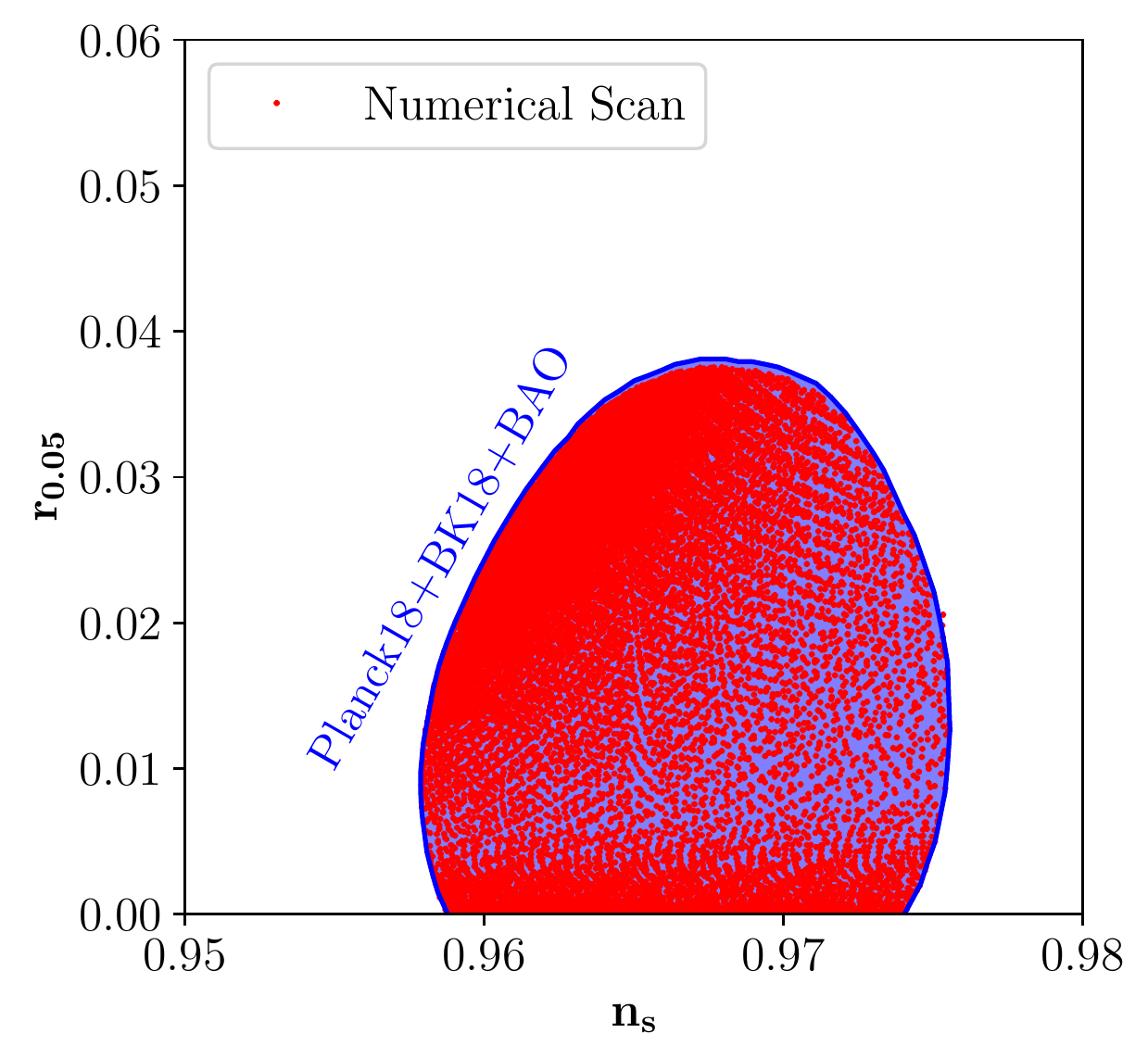}
		\caption {The numerical scan over the free parameters
			$\phi_0, \, \phicmb$ and $\beta$, requiring $50 \leq \ncmb \leq 65$,
			fills the entire allowed region of the $(n_s, \, r)$ plane.}
		\label{fig:rnsscan}
	\end{figure}
	
	In order to explore the full parameter space, we scan over the
	parameters $\phi_0,\, \phicmb$ and $\beta$; $d$ has again been fixed
	such that $\mathcal{P}_{\zeta} \simeq 2.1 \cdot 10^{-9}$. We accept
	all combinations of parameters that yield
	$50 \leq N_{\rm CMB} \leq 65$ and satisfy the constraints on $n_s$ and
	$r$ shown in Fig.~\ref{fig:r-ns}. We see from Fig.~\ref{fig:rnsscan}
	that this scan in fact fills the entire presently allowed region of
	the $(n_s, \, r)$ plane.
	
	\begin{figure}[ht!]
		\centering
		\includegraphics[scale=0.6]{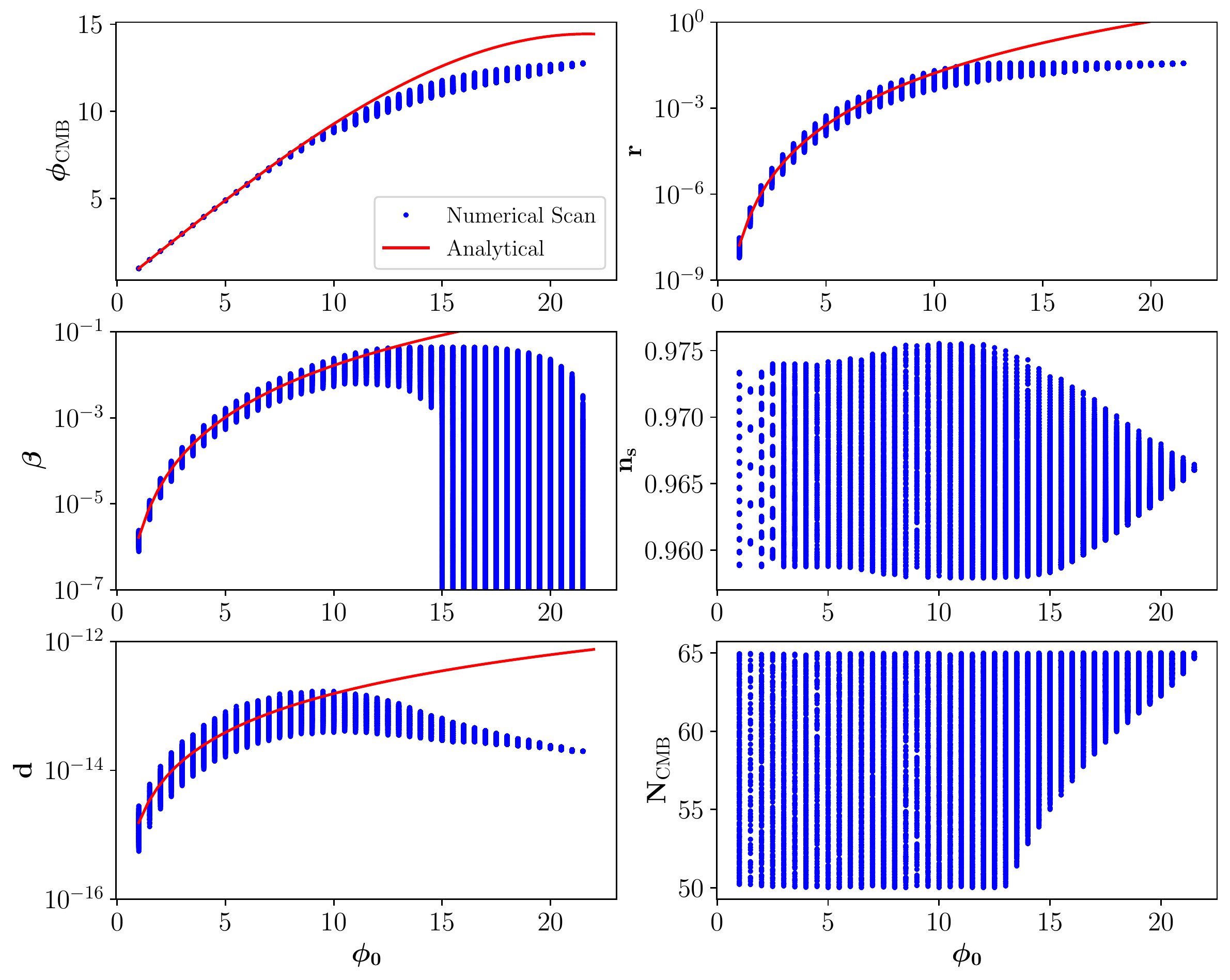}
		\caption {Blue dots represent allowed sets of model parameters (left
			frames) and the corresponding predictions (right frames) using the full
			expressions. The red lines depict the analytical approximation valid
			for $\phi_0 - \phicmb \ll \phi_0$, for fixed $n_s = 0.9659$ and
			$\ncmb = 55$, as described in the text.}
		\label{fig:scan}
	\end{figure}
	The allowed ranges of the model parameters and the resulting
	predictions for $r, \, n_s$ and $\ncmb$ are further shown as function
	of $\phi_0$ in Fig.~\ref{fig:scan}.
	
	The upper left panel shows $\phicmb$. For $\phi_0 \lsim 5$,
	$\phicmb \simeq \phi_0$ can be analytically obtained by \cite{Drees:2021wgd}
	\begin{align}
	\phicmb = \phi_0 (1-\delta_{\rm CMB}) \simeq
	\phi_0 \left(1-7.10\cdot 10^{-4}\phi_0^2\right)\,;
	\end{align}
	here we have used
	$\delta_{\rm CMB}\simeq 1-\frac{\phi_0^2}{48}(1-n_s)$
	(cf. Eq.\eqref{ns}) with $n_s =0.9659$. This analytical approximation
	is shown by the red line, which describes the numerical results very
	well for small $\phi_0$. As we already saw in Fig.~\ref{fig:sample},
	this approximation breaks down for $\phi_0 > 5$, where the difference
	between $\phi_0$ and $\phicmb$ increases quickly. Eventually $\phicmb$
	becomes nearly independent of $\phi_0$, taking values around $12$;
	larger $\phicmb$ would require larger $\epsilon_V$ in order to keep
	$\ncmb$ within the acceptable range, in conflict with the upper bound
	on $r$.
	
	The middle left panel of Fig.~\ref{fig:scan} gives $\beta$ as function
	of $\phi_0$. Here the analytical small $\phi_0$ approximation gives
	\cite{Drees:2021wgd}
	\begin{align}
	\beta \simeq 1.65 \cdot 10^{-6} \phi_0^4\,,
	\end{align}
	for our choices $N_{\rm CMB} =55$ and $n_s =0.9659$; again, this
	reproduces the numerical results for $\phi_0 \lsim 5$. On the other
	hand, for $\phi_0 \geq 15$, a relatively wide range of values of
	$\beta$ is allowed, since for $\beta \ll 1$ the potential at
	$\phi \lsim \phicmb$ only weakly depends on $\beta$; recall from the
	top--left frame that $\phicmb \leq 0.7 \phi_0$ lies well below the
	near--inflection point for these large values of $\phi_0$.
	
	The lower left panel of Fig.~\ref{fig:scan} shows $d$ as
	function of $\phi_0$. In this case the small $\phi_0$ approximation
	yields \cite{Drees:2021wgd}
	\begin{align}
	d \simeq 1.55 \cdot 10^{-15} \phi_0^2\,,
	\end{align}
	for $\mathcal{P}_{\zeta} = 2.1 \cdot 10^{-9}$ and our default values
	of $n_s$ and $\ncmb$.\footnote{The larger scatter of the blue points
		around the red curve in this frame, compared to the top left and
		middle left frames, is largely a plotting artifact. The $y-$axis in
		the latter spans 6 orders of magnitude compared to ``only'' 4 orders
		of magnitude in the lower left frame, making the blue ``bars''
		appear correspondingly shorter. Moreover, for $\phi_0 \lsim 5$ the
		dynamics is more usefully described by the scaled difference
		$\delta$ introduced in eq.(\ref{delta}); the scatter in
		$\delta_{\rm CMB}$ for $\phi_0 \lsim 5$ is similar to that in
		$\beta$ and $d$.} On the other hand, we already saw in the
	discussion of Table~\ref{table:1} that for $\phi_0 > 10$, a smaller
	coupling $d$ is required in order to obtain the correct power
	spectrum.
	
	The upper right panel of Fig.~\ref{fig:scan} depicts $r$ as function
	of $\phi_0$. For $\phi_0 \lsim 5$ this is again quite well described
	by the analytical approximation \cite{Drees:2021wgd}
	\begin{align}
	r\simeq 1.66 \cdot 10^{-8} \phi_0^6\,.
	\end{align}
	For $\phi_0 > 5$ $r$ increases slightly less quickly with increasing
	$\phi_0$; nevertheless for $\phi_0 \gsim 12$ solutions can be found
	that saturate the current upper bound on $r$. In fact, for
	$\phi_0 > 22$ {\em all} solutions that give $\ncmb \leq 65$ predict too
	large a value of $r$.
	
	The middle right panel of Fig.~\ref{fig:scan} gives $n_s$ as function
	of $\phi_0$. For $\phi_0 \leq 13$ essentially the entire currently
	allowed range can be covered by our model. For larger values of $\phi_0$
	the parameter space begins to get squeezed by the conflicting constraints
	$\ncmb \leq 65$ and $r \leq 0.035$. Probably coincidentally the prediction
	of $n_s$ for the largest allowed $\phi_0$ is quite close to the present
	central value.
	
	Finally, the lower right panel of Fig.~\ref{fig:scan} gives results
	for $N_{\rm CMB}$.  Again for $\phi_0 \leq 13$ all values between our
	chosen limits of $50$ and $65$ can be reproduced. For larger $\phi_0$
	the upper bound on $r$ leads to a lower bound on $\ncmb$, since a
	flatter potential reduces the former but increases the latter. As we
	saw above, the two constraints become incompatible for $\phi_0 > 22$.
	
	Let us end this section by summarizing the parameter space for the
	polynomial inflation model (\ref{inflaton_potential2}):
	\begin{align} \label{parameter_space}
	6 \cdot 10^{-16} \lesssim d  \lesssim 2 \cdot 10^{-13}\,;\ \ \
	0 <   \beta  \lesssim 4  \cdot 10^{-2} \,; \ \ \ \phi_0 \lesssim 21.5\,.
	\end{align}
	The upper bound on $\phi_0$ immediately yields a lower bound on the
	cubic potential parameter, $A \gtrsim -57$. Recall that these limits
	have been derived from the observed value of $n_s$, the constraint on
	$\ncmb$, and the upper bound on $r$. It might be worth mentioning that
	in our model the constraint on $n_s$ suffices to derive a theoretical
	upper bound on the tensor--to--scalar ratio, $r \lesssim 0.06$, if
	$\ncmb \gsim 60$; however, for $\ncmb \lsim 55$ significantly larger
	values of $r$ could have been reproduced. The recent bound on $r$
	therefore significantly reduces the allowed parameter space of our
	model. The above discussion also shows that a tighter upper bound on
	$r$ would further limit the parameter space; in particular, the upper
	bound on $\phi_0$ would become smaller if next--generation experiments
	fail to detect CMB tensor modes.
	
	\section{Radiative Stability and Reheating}
	\label{radiative}
	
	So far our analysis has been based on the tree--level potential
	(\ref{inflaton_potential2}). This is only justified if loop
	corrections to the potential are small. In order to check this one can
	compute the $1-$loop Coleman--Weinberg (CW) corrections to the
	potential \cite{Coleman:1973jx} and require that they are subdominant
	compared to the tree--level potential. Here we follow the procedure
	outlined in Ref.~\cite{Drees:2021wgd}. In particular, we focus on the
	potential at $\phi_0$. Here the first and second derivatives of the
	tree--level potential are suppressed by the small parameter $\beta$,
	which increases the relative importance of loop corrections.
	
	The bound on the inflaton self--coupling can be computed by requiring
	the second derivative of the $1-$loop potential at $\phi = \phi_0$ to
	be comparable to the tree--level value. This gives
	\begin{align} \label{dbound}
	\Bigl\lvert \frac{d^2  \ln(16 d \beta)}{\pi^2} \Bigr\rvert < 8 d \beta\,.
	\end{align}
	Since for $\phi_0 \geq 1$ we typically have $\beta \gsim 10^{-6}$
	while $d \lsim 10^{-13}$, this inequality is nearly always
	satisfied. As shown in Fig.~\ref{fig:scan}, for $\phi_0 \gtrsim 14.5$,
	$\beta$ can in principle become arbitrarily small. However, the
	inflationary parameters basically do not change when $\beta$ is
	increased from, say, $10^{-15}$ to a loop corrected value of
	$10^{-13}$; our predictions therefore remain stable in this part of
	parameter space even if the inequality (\ref{dbound}) is violated.
	This therefore does not lead to an additional theoretical constraint on
	the parameter space, in contrast to the small--field version of this
	model \cite{Drees:2021wgd}.
	
	\subsection{Reheating}
	
	In a complete model of very early universe cosmology the inflaton
	field has to couple to external particles for reheating
	\cite{Lyth:2009zz, Kolb:1990vq}. Here we assume $\phi$ couples to
	daughter particles with trilinear couplings in order to fully drain
	the inflaton energy so that a radiation dominated epoch is reproduced
	after reheating. For a scalar field $\phi^{\prime}$, e.g. the standard
	model Higgs field, we introduce a term $g \phi |\phi^{\prime}|^2$ in
	the Lagrangian; for a fermionic field $\chi$, e.g. right--handed
	neutrino, the corresponding term is
	$y \phi \bar{\chi} \chi$.\footnote{The non--perturbative preheating effect in our model is negligible. In  the  bosonic case, i.e. preheating with Higgs production, the trilinear coupling induce tachyonic instabilities \cite{Dufaux:2006ee}, which tends to make preheating efficient. However the Higgs self--coupling gives rise to a positive effective mass $\propto \lambda_{\phi^{\prime}} \langle \phi^{\prime 2}\rangle$ (where $\lambda_{\phi^{\prime}} \sim \mathcal{O}(0.1)$ is the Higgs self-coupling and $\langle \phi^{\prime 2} \rangle$ denotes the variance of the produced Higgs field), which quickly dominates over the (possibly negative) contribution $\propto g \phi$ (where the small trilinear coupling $g$ is bounded by Eq.~\eqref{gbound}), thereby blocking further non--perturbative $\phi \rightarrow \phi'$ energy transfer \cite{Drees:2021wgd, Bernal:2021qrl}. In the fermionic case, Pauli blocking implies that only a small fraction of the energy stored in the inflaton field can be non--perturbatively transferred to $\chi$ particles, unless the daughter particles decay very fast \cite{Allahverdi:2011aj}.} The radiative stability conditions
	\cite{Drees:2021wgd} then read:
	\begin{align}
	\Bigl\lvert \frac{y^4 - 3y^4 \ln (y^2) } {4\pi^2} \Bigr\rvert < 16 d \beta\,;\\
	\frac {1} {8 \pi^2} \left(\frac{g}{\phi_0}\right)^2
	\Bigl\lvert \ln\left(\frac{g}{\phi_0}\right) - 1
	\Bigr\lvert < 8d\beta\,.
	\end{align}
	For typical value of $d \sim 10^{-14}$ and
	$\beta \lesssim \beta_{\text{max}} \simeq 4\cdot 10^{-2}$ (cf.
	Eq.~\eqref{parameter_space}), one has
	\begin{align}
	y &\lesssim y_{\text{max}}\simeq 2.7\cdot 10^{-4}\,; \label{ybound}\\
	\left(\frac{g}{\phi_0}\right) &\lesssim
	\left(\frac{g}{\phi_0}\right)_{\text{max}}\simeq 1.2 \cdot 10^{-7}\,.
	\label{gbound}
	\end{align}

	These upper bounds on the inflaton couplings immediately lead to
	upper bounds on the corresponding partial widths for $\phi \rightarrow
	\phi' \phi'$ and $\phi \rightarrow \bar \chi \chi$ decays, which in
	turn imply upper bounds on the post--inflationary
	reheating temperature. In the instantaneous decay approximation the
	latter is given by \cite{Kolb:1990vq}
	\begin{align} \label{tre}
	\Trh \simeq 1.41 g_{\star}^{-1/4} \Gamma_\phi^{1/2}\,.
	\end{align}
	For the fermionic reheating channel, this becomes
	\begin{align} \label{tref}
	\Trh^\chi \simeq 1.41 g_{\star}^{-1/4} \left(2 \phi_0 \frac{y^2}{8 \pi} \sqrt{d}
	\right)^{1/2} \lesssim 1.1 \cdot 10^{11}~\text{GeV}\,,
	\end{align}
	where we have considered $\phi_0 \lsim 20$, $g_{\star} =106.75$ and used
	Eq.~\eqref{ybound} for the upper bound on $y$. For bosonic reheating,
	the analogous calculation allows even higher reheat temperatures,
	\begin{align} \label{treb}
	\Trh^{\phi'} \simeq 1.41 g_{\star}^{-1/4} \left(\frac{g^2}{8 \pi \,2 \phi_0
		\sqrt{d}} \right)^{1/2}\, \lesssim 2.5\cdot 10^{14}~\text{GeV}\,,
	\end{align}
	where the maximum value $g_{\text{max}}$ reported in
	Eq.~\eqref{gbound} has been utilized.
	
	Both bounds are saturated at the largest allowed value of $\phi_0$,
	where the physical mass of the inflaton
	$m_\phi = 2 \sqrt{d} \phi_0 \simeq 10^{13}$ GeV. This is comfortably
	above the bound (\ref{tref}), which can thus be saturated using simple
	perturbative $\phi \rightarrow \chi \bar\chi$ decays. On the other
	hand, $m_\phi$ is an order of magnitude below the bound
	(\ref{treb}). This bound can therefore only be saturated if one can
	turn an ensemble of $\phi'$ particles with energy
	$\simeq m_\phi/2 \sim 5 \cdot 10^{12}$ GeV into a thermal bath with
	much higher temperature, which can only happen via scattering
	reactions that {\em reduce} the number of particles, in particular
	$3 \rightarrow 2$ scattering reactions. It is not clear whether the
	rate of such reactions is sufficiently high that the bound
	(\ref{treb}) can be saturated; in the absence of such reactions it
	would have to be replaced by $\Trh^{\phi'} \lsim m_\phi/2$. The same
	remark holds for the maximal temperature of the radiation bath, which
	is typically attained well before reheating is completed and can be
	significantly larger than $\Trh$ \cite{Kolb:1990vq}: in the absence of
	fast $3 \rightarrow 2$ reactions it is also bounded by $m_\phi/2$.
	
	
	\section{Eternal Polynomial Inflation}
	\label{eternal}
	
	In the previous sections, we have worked out the parameter space
	consistent with Planck 2018 \eqref{planck2018} and BICEP/Keck 2018
	\eqref{BK2018} and investigated the radiative stability of the
	inflaton potential as well as perturbative reheating for the
	polynomial inflation model. All this happened at field values (well)
	below $\phi_0$. In this section we analyze the situation at
	$\phi \gsim \phi_0$.  In particular, we are interested in the energy
	scale(s) at which ``eternal'' inflation could have occurred within the
	allowed parameter space.
	
	
	\subsection{Eternal Phase I}
	
	During the SR phase, one can neglect the acceleration term in
	eq.~(\ref{eom1}), so that the classical inflaton field evolves as
	\begin{align} \label{eom2}
	\dot{\phi} \approx -\frac{V^{\prime}(\phi)}{3 H } \,.
	\end{align}
	This predicts a classical field excursion per Hubble time
	\begin{align}\label{per_hubble}
	\Delta \phi_{\rm cl} = \frac{ |\dot{\phi}|}{ H } \approx
	\frac{|V^{\prime}(\phi)|}{3 H^2 } \approx  \frac{|V^{\prime}(\phi)|}{V }
	= \sqrt{2\epsilon_V}\,.
	\end{align}
	On the other hand, since the inflaton field is very weakly coupled its
	quantum fluctuations follow a Gaussian probability distribution. In
	the quasi de Sitter background during SR inflation, the typical size
	of these quantum fluctuation over one Hubble time is given by
	$\delta \phi_{\rm qu} = \frac{H}{2\pi}$ \cite{Starobinsky:1982ee,
		Linde:1982uu}.
	
	Assume that the quantum fluctuation dominates over the classical field
	excursion, i.e. $\delta \phi_{\rm qu} > \Delta \phi_{\rm cl}$. In this
	case the inflaton field in a given Hubble volume is almost as likely
	to move ``uphill'', towards larger values, as it is to move towards
	the minimum of the potential. Since after a small number of e--folds
	of inflation the inflaton field fills a great many Hubble volumes
	which henceforth evolve independently, it is virtually guaranteed that
	in some of these volumes the field does indeed move uphill. Since
	these regions subsequently will expand faster (owing to the larger
	Hubble parameter), in this picture ``most'' of space will continue
	inflating forever, even though inflation clearly must have ended in
	our own Hubble patch. This is known as eternal inflation.
	
	From eq.(\ref{per_hubble}) and
	$\delta \phi_{\rm qu} = \frac{H}{2\pi}$, the condition for eternal
	inflation is
	\begin{align} \label{condition}
	\frac{H}{2\pi} > \sqrt{2\epsilon_V}
	\Leftrightarrow \frac{H^2}{8\pi^2\epsilon_V} >1\,.
	\end{align}
	Hence the condition for eternal inflation is satisfied if the leading
	order prediction of the amplitude of curvature perturbations
	$\mathcal{P}_{\zeta}$ exceeds unity \cite{Barenboim:2016mmw}.
	
	As already emphasized in sec.~\ref{analytical}, for $\phi \gg \phi_0$
	effectively our model behaves like quartic inflation, i.e. the
	inflaton potential can be simplified to $V = d\, \phi^4$, so that
	\begin{align} \label{hubble}
	H = \sqrt{\frac{d}{3}} \phi^2\,,
	\end{align}
	and 
	\begin{align} \label{pow}
	\frac{H^2}{8\pi^2\epsilon_V}  = \frac{d \phi^6} {192 \pi ^2}\,.
	\end{align}
	From condition~\eqref{condition} eternal inflation then
	requires 
	\begin{align}
	\phi^2 > \left(\frac{192 \pi ^2}{d}\right)^{1/3}\,.
	\end{align}
	This in turn leads to a lower bound on the Hubble parameter:
	\begin{align} \label{threshold}
	H > H^{\rm c}_{\rm EI} = 4 \pi^{2/3} \left(\frac{d}{3}\right)^{1/6}\,,
	\end{align}
	which only depends on $d$. Once $H > H^{\rm c}_{\rm EI} $, an eternal
	inflationary phase can occur.
	
	The usual monomial chaotic $\lambda \phi^4$ inflation model requires
	$\lambda \sim 10^{-12}$ \cite{Guth:2007ng} in order to match the
	normalization of the power spectrum; eq.~(\ref{threshold}) then yields
	$H^{\rm c \prime}_{\rm EI}\approx 0.07$. In the last section we saw
	that in our polynomial scenario the quartic coupling $d$ needs to be
	somewhat smaller,
	$6 \cdot 10^{-16} \lesssim d \lesssim 2 \cdot 10^{-13}$. The
	corresponding threshold value of the inflaton field is (in Planckian
	units):
	\begin{align} \label{phicrit}
	460 \lesssim 	\phi_{\rm c} 	 \lesssim  1211\,,
	\end{align}
	which evidently is indeed well above $\phi_0$. One can further work
	out the threshold of the corresponding inflationary scale (again in
	Planckian units):
	\begin{align} \label{th1}
	0.02 \lesssim 	H^{\rm c}_{\rm EI} \lesssim  0.05\,.
	\end{align}
	Evidently the corresponding energy scale is well below the Planck scale (and
	also somewhat below the scale required for eternal inflation in pure quartic
	inflation); hence our semi--classical treatment, which ignores ``quantum
	gravity'' effects, may be valid.  
	
	We thus conclude that our model does allow for an epoch of ``eternal''
	while reproducing all present measurements of inflationary parameters.
	
	
	\subsection{Eternal Phase II}
	
	The discussion in the previous section shows that eternal inflation
	should occur in the polynomial model if the inflaton field and the
	Hubble parameter ever exceeded the critical values (\ref{phicrit}) and
	(\ref{th1}), respectively. Here we show that, at least in part of the
	allowed parameter space, a second, later epoch of eternal inflation
	will occur.
	
	We have seen in Fig.~\ref{fig:scan} that the parameter $\beta$ can be
	arbitrarily small if $\phi_0 \gsim 15$. Recall that a very small
	$\beta$ implies that the potential at $\phi_0$ is very flat. Hence it
	is expected that a (second) eternal phase can occur. Since this
	eternal phase appears when the inflaton is near the saddle point
	$\phi_0$, which is much smaller than $\phi_{\rm c}$ analyzed above,
	one can expect that the corresponding Hubble scale should be much
	lower than $H^{\rm c}_{\rm EI}$ of eq.(\ref{th1}).\footnote{This has
		some similarity to eternal hilltop inflation investigated in
		\cite{Barenboim:2016mmw}. However, in our case the inflaton first
		rolls down to a plateau around the saddle point; in hilltop
		inflation, one has to impose as initial condition that the inflaton
		starts near a (local) maximum of the potential, which can also be
		very flat.}
	
	In order to obtain the maximum value of $\beta$ that allows a second
	phase of eternal inflation, we again use condition
	Eq.~\eqref{condition}, identical to the leading order prediction of
	the power spectrum being larger than unity as mentioned earlier. Since
	the potential is flattest at $\phi = \phi_0$, we can use
	eq.(\ref{powerspectrum2}) with $\delta = 0$ to derive the condition for
	the existence of a second epoch of eternal inflation:
	\begin{equation} \label{eq:second}
	\beta < \frac {\sqrt{d} \phi_0^3} {144 \pi} \simeq 10^{-6}\,;
	\end{equation}
	For the numerical value we have used $d \sim 10^{-14}$ and
	$\phi_0 \sim 20$.
	
	The potential at $\phi_0$ is given by $d \phi_0^4/3$, up to
	corrections of relative order $\beta$ which are evidently completely
	negligible here. This epoch of eternal inflation would thus have a
	Hubble parameter
	\begin{align} \label{saddle_scale}
	H_{\rm EI}  = \frac{\sqrt{d}}{3 }  \phi_0^2 \sim 10^{-5}\,;
	\end{align}
	this is at least three orders of magnitude smaller than the one given in
	eq.(\ref{th1}).
	
	The width of this second region in field space that allows eternal
	inflation is also important. Again from eq.(\ref{powerspectrum2}) we
	see that the evolution of the inflaton field will be dominated by
	quantum fluctuations as long as
	\begin{equation} \label{eq:delta_c}
	\delta < \delta_{\rm c} = \frac { d^{1/4} \phi_0^{3/2}} {6 \sqrt{2\pi}}\,.
	\end{equation}
	In units of the size $H_{\rm EI}/(2\pi)$ of random walk steps the half width
	of the region in field space allowing eternal inflation is thus
	\begin{equation} \label{eq:width}
	\frac{ \delta_{\rm c} \phi_0} { H_{\rm EI}/(2\pi)}
	= \sqrt{ \frac{\pi}{2} } \frac { \phi_0^{1/2} } {d ^{1/4}}
	\simeq 1.8 \cdot 10^4\,.
	\end{equation}
	Given that the average excursion in a random walk is proportional to
	the square of the number of steps taken times the (typical) step size,
	starting from $\phi = \phi_0$ the field would thus typically need more
	than $10^8$ steps to leave the region where quantum fluctuations
	dominate the dynamics. Moreover, half the time the random walk would
	end at $\phi > \phi_0 \left(1 + \delta^{\rm c} \right)$, in which case
	the classical field evolution would bring the field back into the
	range where quantum effects dominate. These arguments indicate that in
	our model indeed ``most of'' space would inflate eternally if $\phi$
	ever reached the region very close to $\phi_0$ and $\beta < 10^{-6}$.
	
	\begin{figure}[ht!]
		\centering
		\includegraphics[scale=0.8]{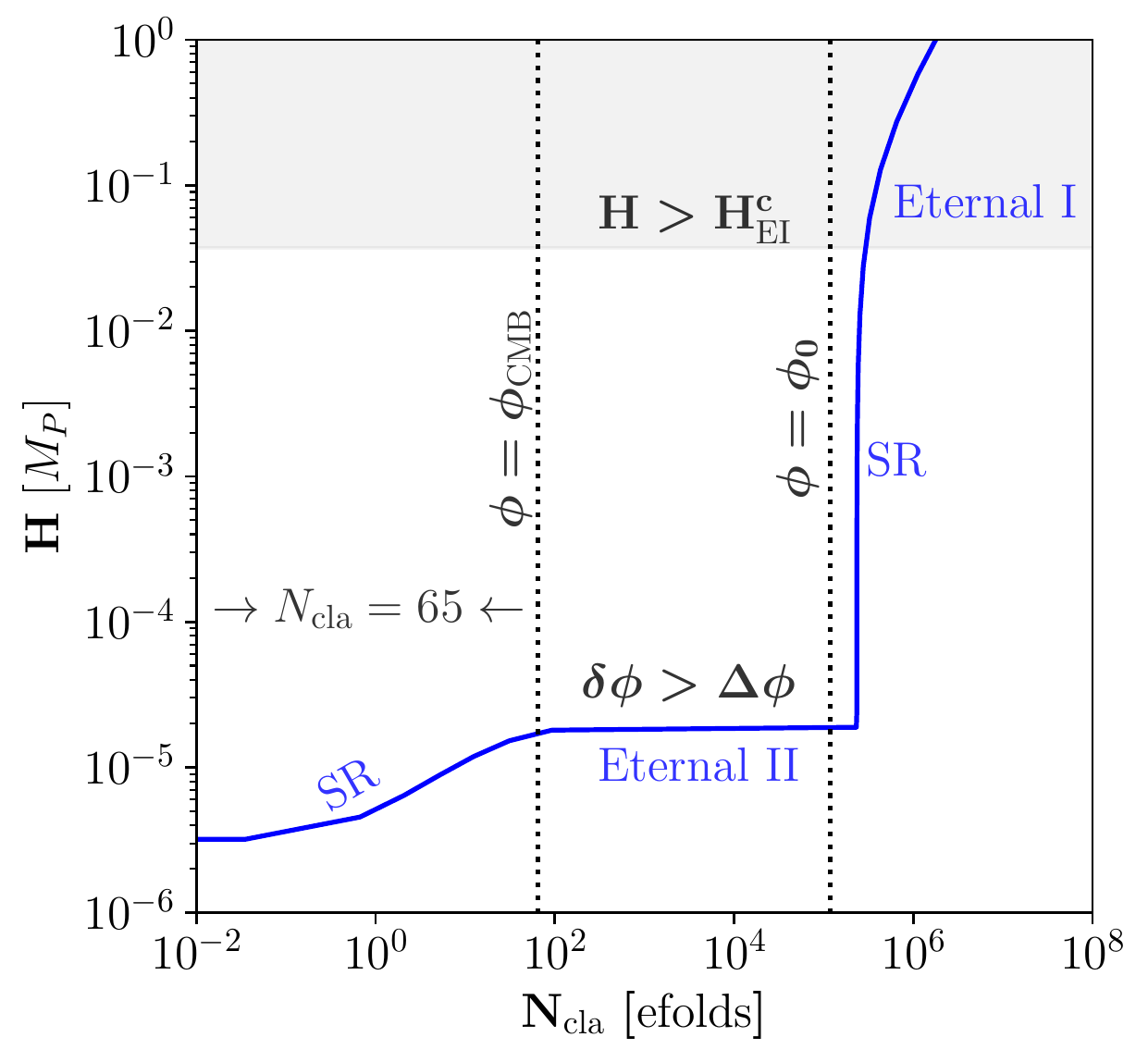}
		\caption {Evolution of the Hubble parameter as function of the
			classical prediction for the number of e--folds $N_{\rm cla}$
			(depending on $\phi$ via Eq.~\eqref{ncmb}), for
			$\phi_0 = 20,~ \beta =10^{-7}$ and $d \simeq 2 \cdot
			10^{-14}$. Inflation can be eternal in the gray shaded region where
			$H>H^{\rm c}_{\rm EI} \simeq 0.04~M_{\rm Pl}$. For
			$H < H^{\rm c}_{\rm EI}$ a period of the usual SR inflation
			follows. For the given choice of a very small $\beta$ a {\em second}
			epoch of eternal inflation occurs (between the dashed vertical
			lines), with Hubble parameter as low as
			$\mathcal{O}(10^{-5})~M_{\rm Pl}$.}
		\label{two_phase}
	\end{figure}

	A typical inflationary trajectory starting at very large field values
	$\phi > \phi_{\rm c}$ of eq.(\ref{phicrit}) is shown in
	Fig.~\ref{two_phase}, which plots the Hubble parameter as function of
	the classical prediction $N_{\rm cla}$ of the number of e--folds that
	occur after the inflaton field had a certain value, given by
	eq.(\ref{ncmb}); larger $N_{\rm cla}$ correspond to larger $\phi$.  Of
	course, for the field ranges allowing eternal inflation, i.e. for
	$\phi > \phi_{\rm c}$ and for
	$\phi \in [\phi_0(1-\delta_{\rm c}), \phi_0(1+\delta_c)]$, the {\em
		actual} number of e--folds by which our Hubble patch expanded was
	likely very much larger than the classical prediction.
	
	For field values below $\phi_{\rm c}$ the Hubble parameter seems to
	drop very steeply. However, if $\phi_0 \gsim 4$ the SR conditions are
	satisfied for all $\phi \gsim \phi_0$; essentially deterministic
	inflation therefore lasts from the end of the first, high--energy
	stage of eternal inflation to the onset of the second epoch of eternal
	inflation where $\phi \simeq \phi_0$. This is in contrast to the
	small--field version of this model, where the SR conditions are
	violated for some range of field values above $\phi_0$, and inflation
	around $\phi_0$ is always deterministic \cite{Drees:2021wgd}. Finally,
	once $\phi < \phi_0 \left( 1 - \delta_{\rm c} \right)$ inflation is
	deterministic again, including the last $\sim 65$ e--folds of
	inflation with $\phi \leq \phicmb$.
	
	Recalling the results of Fig.~\ref{fig:scan}, we conclude that model
	parameters (a subset of \eqref{parameter_space}) with:
	\begin{align} \label{eternalps}
	15 \lesssim \phi_0 \lesssim 21.5\,;\ \ \
	0<  \beta \lesssim \mathcal{O}(10^{-6})\,;\ \ \
	2 \cdot 10^{-14} \lesssim d \lesssim 6 \cdot 10^{-14}\,,
	\end{align}
	satisfy all observational constraints and allow a second eternal
	inflationary phase. Note that the second eternal phase can only occur
	if $\phi_0$ is rather large, which implies $r \sim \mathcal{O}(10^{-2})$.
	The part of parameter space allowing the (unusual) low scale second
	eternal phase can thus be tested by the next generation CMB experiments,
	e.g. CORE \cite{Bouchet:2011ck}, AliCPT \cite{Li:2017drr},
	LiteBIRD \cite{Matsumura:2013aja} and CMB-S4 \cite{Abazajian:2019eic},
	with expected sensitivity down to $r \sim \mathcal{O}(10^{-3})$.

	\section{Summary and Conclusions}
	\label{sumup}
	
	In this paper we revisited large field inflation with a single
	inflaton. We investigated a model where the inflaton potential is a
	polynomial of degree four. Current observations then require that the
	potential features a near saddle point at $\phi_0$, making the
	potential concave at $\phi \lsim \phicmb < \phi_0$ as required by the
	Planck 2018 data.
	
	The model was described in Sec.~\ref{setup}. The potential contains
	three free parameters: an overall (quartic) coupling strength $d$; the
	location $\phi_0$ of the almost saddle point; and $\beta \ll 1$ to
	determine the deviation from a true saddle point, with smaller $\beta$
	making the potential flatter for $\phi \sim \phi_0$. The parameters
	$\phi_0$ and $\beta$ thus determine the shape of the potential, while
	the overall normalization, given by $d$, can be fixed from the
	normalization of the power spectrum of curvature perturbation
	$\mathcal{P}_{\zeta}$. The value $\phicmb$ of the inflaton field when
	the cosmic microwave background (CMB) scales first left the horizon is
	another important free parameter. It allows us to choose $\phi_0$
	within a rather wide range, and then use the measured spectral index
	$n_s$ as well as the number $\ncmb$ of e--folds of inflation generated
	after CMB scales first left the horizon to determine, or constrain,
	$\phicmb$ and $\beta$. We also discuss various analytical
	approximations. In particular, for $\phi_0 \lsim 5$ (in Planckian
	units, which we use throughout) $\phi_0 - \phicmb \ll \phi_0$ and the
	fully analytical treatment developed for the small field version of
	this model \cite{Drees:2021wgd} still holds to good approximation.
	
	In Sec.~\ref{model_parameters} a full scan of the parameter space
	consistent with most resent Planck and BICEP/Keck 2018 observations at
	the $2\sigma$ level (cf. \eqref{planck2018} and \eqref{BK2018}) is
	described. The final result is summarized in
	(\ref{parameter_space}). The predictions for $r$ range from
	unobservably small, $\mathcal{O}(10^{-8})$, to the current upper
	bound. In fact, the current bound $r \leq 0.035$ together with the
	constraint $\ncmb \leq 65$ leads to the upper bound
	$\phi_0 \lsim 21.5$. Moreover, we predict negative running of the
	spectral index, $\alpha \sim -\mathcal{O}(10^{-3})$, which might be
	testable in the near future \cite{Munoz:2016owz}. A large set of examples are listed in
	Table \ref{table:1}. To our knowledge, this is the first such
	comprehensive scan of parameter space of polynomial inflation taking
	into account the most recent CMB data. Of course, the currently
	allowed region described by eqs.(\ref{parameter_space}), in particular
	the upper bound on $\phi_0$, should be further constrained once more
	precise CMB experiments are performed, such as CORE
	\cite{Bouchet:2011ck}, AliCPT \cite{Li:2017drr}, LiteBIRD
	\cite{Matsumura:2013aja} and CMB-S4 \cite{Abazajian:2019eic}, which
	could probe all $r \gsim \mathcal{O}(10^{-3})$.
	
	In Sec.~\ref{radiative} we showed that radiative stability of the
	inflaton potential near the inflection-point $\phi_0$ leads to
	relatively mild constraints for our large field model, in sharp
	contrast to its small field version \cite{Drees:2021wgd}. In
	particular, the one--loop Coleman--Weinberg (CW) corrections to the
	potential due to the self--interactions of the inflaton are always
	harmless. Moreover, large reheat temperatures, up to $10^{11}$
	$(2 \cdot 10^{14})$ GeV are in principle possible for perturbative
	inflaton decay into fermionic (bosonic) final states. However, we
	remind the reader that temperatures above half the inflaton mass
	$m_\phi$ can only be reached if the rate for reactions that reduce the
	number of particles (e.g. $3 \rightarrow 2$) is sufficiently high;
	note that $m_\phi \leq 10^{13}$ GeV in this model.
	
	In Sec.~\ref{eternal} the possibility of eternal inflation is
	discussed. This is generally expected to occur in models of polynomial
	inflation, assuming the field value or, equivalently, the inflationary
	Hubble parameter reached sufficiently high values. In our case the
	critical Hubble parameter only depends on the size of the quartic
	coupling $d$, and lies in the range
	$0.02 \lesssim H^{\rm c}_{\rm EI} \lesssim 0.05$; this is somewhat
	below the value of $0.07$ needed in the monomial $\lambda \phi^4$
	model (which is in any case excluded by the upper bound on $r$).
	
	More interestingly, we find that there exists another possibility to
	realize eternal inflation in our scenario. This occurs when the
	potential is very flat around $\phi_0$: for $\beta \lsim 10^{-6}$
	quantum fluctuations can dominate over the classical evolution already
	at $\phi \simeq \phi_0$, compared to $\phi \gsim {\cal O}(10^3)$ in
	the first epoch of eternal inflation. This second epoch of eternal
	inflation features a much smaller Hubble parameter,
	$H \sim \mathcal{O}(10^{-5})$, which is of the same order of magnitude
	as that when the CMB pivot scale $k_{\star} = 0.05\ \rm{Mpc}^{-1}$
	first crossed out of the horizon. Nevertheless there will be many
	e--folds of deterministic inflation between the end of the second
	epoch of eternal inflation and the era when the CMB scales first
	crossed out of the horizon. We are therefore not aware of any
	immediate observational consequences of this ``late'' epoch of eternal
	inflation. However, since $\beta \lsim 10^{-6}$ is possible only for
	$\phi_0 \gsim 15$, which in turn implies $r \gsim 0.01$, at least in
	our model this possibility can be tested by the next round of CMB
	experiments.
	
	This novel scenario featuring two epochs of eternal inflation, as
	depicted in fig.~\ref{two_phase}, might also be conceptually
	interesting. Eternal inflation is the only known mechanism that might
	be able to populate the ``landscape'' of superstring theory
	\cite{Susskind:2003kw, Clifton:2007en, Linde:2006nw}. During eternal
	inflation not only the inflaton field undergoes a random walk in field
	space, but so does {\em every} field whose mass is below the
	inflationary Hubble parameter. This might include many of the scalar
	fields (from the four--dimensional perspective) that determine the
	sizes of physical couplings in string theory. During the first, high
	scale epoch of eternal inflation these fields will be sampled with a
	typical step size $H/(2\pi) \gsim 5\cdot 10^{-3}$. Since the second
	epoch of eternal inflation has a thousand times smaller Hubble
	parameter, the step size of the random walk in field space is also
	thousand times smaller during this second epoch. This might allow to
	much more efficiently ``home in'' on relatively small features of the
	landscape.
	
	In summary, we have presented a successful large field polynomial
	model, worked out the complete allowed parameter space
	(\ref{parameter_space}) and offered the corresponding inflationary
	predictions. Combined with the earlier analysis of the small field
	version of this model \cite{Drees:2021wgd} this offers the most
	complete analysis of the polynomial inflation model after Planck and
	BICEP/Keck 2018. We also pointed out for the first time that in our
	model the early history of the universe might feature {\em two} epochs
	of eternal inflation, at quite different energy scales.
	
	\clearpage
	\appendix
	\section{General Expression for the Number of e-Folds}
	\begin{align}
	N_{\rm CMB} &= \int^{\phicmb}_{\phiend} \frac{1}{\sqrt{2 \epsilon_V}} d\phi
	\nonumber \\
	& = \int^{\phicmb}_{\phiend} \frac{\phi}{2}\left[ \frac{32\phi^2
		-32A\left(\beta-1\right)\phi  +9A^2}{64\phi^2 -48A\left(\beta-1\right)
		\phi  +9A^2 }  \right]d\phi\,
	\nonumber \\
	& = \int^{\phicmb}_{\phiend} \frac{\phi}{2}\left[ 1+ \frac{-32\phi^2
		+ 16A\left(\beta-1\right)\phi}{64\phi^2 -48A\left(\beta-1\right)\phi
		+ 9A^2 }  \right]d\phi\,
	\nonumber \\
	& = \int^{\phicmb}_{\phiend} \frac{\phi}{2}\left \{ 1+ \frac{-32\phi^2
		+16A\left(\beta-1\right)\phi}{ \left[8\phi -3A(\beta - 1) \right]^2
		- 9A^2(\beta^2 -2 \beta)}  \right \}d\phi\,
	\nonumber \\
	& = \int^{\phicmb}_{\phiend} \frac{\phi}{2}\left \{ 1+ \frac{-32\phi^2
		+ 16A \left(\beta-1\right)\phi}{ 64\left[\phi -3/8A(\beta - 1) \right]^2
		- 9A^2(\beta^2 -2 \beta)}  \right \}d\phi\,
	\\
	& = \int^{\phicmb}_{\phiend} \frac{\phi}{2}\left \{ 1-\frac{1}{2} \frac{\phi^2 }
	{\left[\phi +\phi_0(\beta - 1) \right]^2 -\phi_0^2(\beta^2 -2 \beta)}
	-\frac{2}{3}\frac{ \phi_0\left(\beta-1\right)\phi}
	{\left[\phi +\phi_0(\beta - 1) \right]^2
		-\phi_0^2(\beta^2 -2 \beta)}  \right \}d\phi\,.
	\nonumber
	\end{align}
	The three integrals can be evaluated analytically:
	\begin{equation}
	\int d \phi \frac{\phi}{2}  = \frac{\phi^2}{4}\,;
	\end{equation}
	\begin{align}
	\int  -\frac{1}{4} \frac{\phi^3}{\left[\phi -m \right]^2 +n^2}\, d\phi
	= \frac{1}{8} & \Bigg\{ 5m^2 -4 m\, \phi - \phi^2 +
	\frac{2(m^3 -3m\,n^2)\,}{n}
	\arctan\left[\frac{ m-\phi }{n}\right]
	\nonumber  \\  &
	-(3m^2 - n^2)
	\log \left[n^2+(m-\phi)^2\right]\Bigg\}\,;
	\end{align}
	and
	\begin{align}
	\int & \frac{1}{3} \frac{d\phi \,   m\, \phi^2}{\left[\phi -m \right]^2 +n^2}
	= \frac{m}{3} \left\{ \phi -\frac{(m^2- n^2)}{n} \arctan
	\left[\frac{ m-\phi }{n}\right]  + m \log \left[n^2+(m-\phi)^2\right]
	\right\}\,,
	\end{align}
	with $m = \phi_0(1-\beta)$ and $n^2 = \phi_0^2(2\beta-\beta^2)$.
	Combining these results, we obtain
	\begin{align}
	\ncmb &=\left\{  -\frac{ m^3 + 5mn^2 }{12n} \arctan
	\left( \frac{ m-\phi }{n} \right)  + \frac{5 m^2}{8} - \frac{m \phi}{6}
	+ \frac {\phi^2}{8}
	\nonumber \right. \\  &\left. \hspace*{7mm}
	-\frac{(m^2 - 3n^2)} {24} \ln\left[n^2 + (m-\phi)^2\right] \right\}
	\Bigg \vert^{\phicmb}_{\phiend}
	\\
	&\simeq \frac{1}{24} \left \{ 3\phi^2 - 4\phi \phi_0  + 15\phi_0^2
	-\phi_0^2\sqrt{\frac{2}{\beta}} \arctan \left( \frac{\phi_0 -\phi }
	{\sqrt{2\beta}\phi_0} \right)
	- \phi_0^2\ln\left[(\phi_0 - \phi)^2 \right] \right \}
	\Bigg \vert^{\phicmb}_{\phiend}\,.
	\nonumber
	\end{align}
	In the last step we have assumed $\beta \ll 1$, which is true in the
	allowed parameter space, but cannot be assumed a priori.
	
	\bibliographystyle{JHEP}
	\bibliography{biblio}

\providecommand{\href}[2]{#2}\begingroup\raggedright\begin{thebibliography}{10}

\bibitem{Starobinsky:1980te}
A.A.~Starobinsky, \emph{{A New Type of Isotropic Cosmological Models Without
  Singularity}},
  \href{https://doi.org/10.1016/0370-2693(80)90670-X}{\emph{Phys. Lett.}
  {\bfseries 91B} (1980) 99}.

\bibitem{Guth:1980zm}
A.H.~Guth, \emph{{The Inflationary Universe: A Possible Solution to the Horizon
  and Flatness Problems}},
  \href{https://doi.org/10.1103/PhysRevD.23.347}{\emph{Phys. Rev. D} {\bfseries
  23} (1981) 347}.

\bibitem{Linde:1981mu}
A.D.~Linde, \emph{{A New Inflationary Universe Scenario: A Possible Solution of
  the Horizon, Flatness, Homogeneity, Isotropy and Primordial Monopole
  Problems}}, \href{https://doi.org/10.1016/0370-2693(82)91219-9}{\emph{Phys.
  Lett. B} {\bfseries 108} (1982) 389}.

\bibitem{Albrecht:1982wi}
A.~Albrecht and P.J.~Steinhardt, \emph{{Cosmology for Grand Unified Theories
  with Radiatively Induced Symmetry Breaking}},
  \href{https://doi.org/10.1103/PhysRevLett.48.1220}{\emph{Phys. Rev. Lett.}
  {\bfseries 48} (1982) 1220}.

\bibitem{Mukhanov:1981xt}
V.F.~Mukhanov and G.V.~Chibisov, \emph{{Quantum Fluctuations and a Nonsingular
  Universe}}, {\emph{JETP Lett.} {\bfseries 33} (1981) 532}.

\bibitem{Martin:2013tda}
J.~Martin, C.~Ringeval and V.~Vennin, \emph{{Encyclop\ae{}dia Inflationaris}},
  \href{https://doi.org/10.1016/j.dark.2014.01.003}{\emph{Phys. Dark Univ.}
  {\bfseries 5-6} (2014) 75} [\href{https://arxiv.org/abs/1303.3787}{{\ttfamily
  1303.3787}}].

\bibitem{Planck:2018vyg}
{\scshape Planck} collaboration, \emph{{Planck 2018 results. VI. Cosmological
  parameters}},
  \href{https://doi.org/10.1051/0004-6361/201833910}{\emph{Astron. Astrophys.}
  {\bfseries 641} (2020) A6}
  [\href{https://arxiv.org/abs/1807.06209}{{\ttfamily 1807.06209}}].

\bibitem{BICEPKeck:2021gln}
{\scshape BICEP/Keck} collaboration, \emph{{Improved Constraints on Primordial
  Gravitational Waves using Planck, WMAP, and BICEP/Keck Observations through
  the 2018 Observing Season}},
  \href{https://doi.org/10.1103/PhysRevLett.127.151301}{\emph{Phys. Rev. Lett.}
  {\bfseries 127} (2021) 151301}
  [\href{https://arxiv.org/abs/2110.00483}{{\ttfamily 2110.00483}}].

\bibitem{Silverstein:2008sg}
E.~Silverstein and A.~Westphal, \emph{{Monodromy in the CMB: Gravity Waves and
  String Inflation}},
  \href{https://doi.org/10.1103/PhysRevD.78.106003}{\emph{Phys. Rev. D}
  {\bfseries 78} (2008) 106003}
  [\href{https://arxiv.org/abs/0803.3085}{{\ttfamily 0803.3085}}].

\bibitem{McAllister:2008hb}
L.~McAllister, E.~Silverstein and A.~Westphal, \emph{{Gravity Waves and Linear
  Inflation from Axion Monodromy}},
  \href{https://doi.org/10.1103/PhysRevD.82.046003}{\emph{Phys. Rev. D}
  {\bfseries 82} (2010) 046003}
  [\href{https://arxiv.org/abs/0808.0706}{{\ttfamily 0808.0706}}].

\bibitem{Hodges:1989dw}
H.M.~Hodges, G.R.~Blumenthal, L.A.~Kofman and J.R.~Primack, \emph{{Nonstandard
  Primordial Fluctuations From a Polynomial Inflaton Potential}},
  \href{https://doi.org/10.1016/0550-3213(90)90177-F}{\emph{Nucl. Phys. B}
  {\bfseries 335} (1990) 197}.

\bibitem{Destri:2007pv}
C.~Destri, H.J.~de~Vega and N.G.~Sanchez, \emph{{MCMC analysis of WMAP3 and
  SDSS data points to broken symmetry inflaton potentials and provides a lower
  bound on the tensor to scalar ratio}},
  \href{https://doi.org/10.1103/PhysRevD.77.043509}{\emph{Phys. Rev. D}
  {\bfseries 77} (2008) 043509}
  [\href{https://arxiv.org/abs/astro-ph/0703417}{{\ttfamily
  astro-ph/0703417}}].

\bibitem{Aslanyan:2015hmi}
G.~Aslanyan, L.C.~Price, J.~Adams, T.~Bringmann, H.A.~Clark, R.~Easther et~al.,
  \emph{{Ultracompact minihalos as probes of inflationary cosmology}},
  \href{https://doi.org/10.1103/PhysRevLett.117.141102}{\emph{Phys. Rev. Lett.}
  {\bfseries 117} (2016) 141102}
  [\href{https://arxiv.org/abs/1512.04597}{{\ttfamily 1512.04597}}].

\bibitem{Allahverdi:2006iq}
R.~Allahverdi, K.~Enqvist, J.~Garcia-Bellido and A.~Mazumdar, \emph{{Gauge
  invariant MSSM inflaton}},
  \href{https://doi.org/10.1103/PhysRevLett.97.191304}{\emph{Phys. Rev. Lett.}
  {\bfseries 97} (2006) 191304}
  [\href{https://arxiv.org/abs/hep-ph/0605035}{{\ttfamily hep-ph/0605035}}].

\bibitem{Nakayama:2013jka}
K.~Nakayama, F.~Takahashi and T.T.~Yanagida, \emph{{Polynomial Chaotic
  Inflation in the Planck Era}},
  \href{https://doi.org/10.1016/j.physletb.2013.06.050}{\emph{Phys. Lett. B}
  {\bfseries 725} (2013) 111}
  [\href{https://arxiv.org/abs/1303.7315}{{\ttfamily 1303.7315}}].

\bibitem{Nakayama:2013txa}
K.~Nakayama, F.~Takahashi and T.T.~Yanagida, \emph{{Polynomial Chaotic
  Inflation in Supergravity}},
  \href{https://doi.org/10.1088/1475-7516/2013/08/038}{\emph{JCAP} {\bfseries
  08} (2013) 038} [\href{https://arxiv.org/abs/1305.5099}{{\ttfamily
  1305.5099}}].

\bibitem{Kallosh:2014xwa}
R.~Kallosh, A.~Linde and A.~Westphal, \emph{{Chaotic Inflation in Supergravity
  after Planck and BICEP2}},
  \href{https://doi.org/10.1103/PhysRevD.90.023534}{\emph{Phys. Rev. D}
  {\bfseries 90} (2014) 023534}
  [\href{https://arxiv.org/abs/1405.0270}{{\ttfamily 1405.0270}}].

\bibitem{Li:2014zfa}
T.~Li, Z.~Sun, C.~Tian and L.~Wu, \emph{{The Renormalizable Three-Term
  Polynomial Inflation with Large Tensor-to-Scalar Ratio}},
  \href{https://doi.org/10.1140/epjc/s10052-015-3508-4}{\emph{Eur. Phys. J. C}
  {\bfseries 75} (2015) 301} [\href{https://arxiv.org/abs/1407.8063}{{\ttfamily
  1407.8063}}].

\bibitem{Gao:2015yha}
T.-J.~Gao and Z.-K.~Guo, \emph{{Inflection point inflation and dark energy in
  supergravity}}, \href{https://doi.org/10.1103/PhysRevD.91.123502}{\emph{Phys.
  Rev. D} {\bfseries 91} (2015) 123502}
  [\href{https://arxiv.org/abs/1503.05643}{{\ttfamily 1503.05643}}].

\bibitem{Musoke:2017frr}
N.~Musoke and R.~Easther, \emph{{Expectations for Inflationary Observables:
  Simple or Natural?}},
  \href{https://doi.org/10.1088/1475-7516/2017/12/032}{\emph{JCAP} {\bfseries
  12} (2017) 032} [\href{https://arxiv.org/abs/1709.01192}{{\ttfamily
  1709.01192}}].

\bibitem{Linde:2007jn}
A.D.~Linde and A.~Westphal, \emph{{Accidental Inflation in String Theory}},
  \href{https://doi.org/10.1088/1475-7516/2008/03/005}{\emph{JCAP} {\bfseries
  03} (2008) 005} [\href{https://arxiv.org/abs/0712.1610}{{\ttfamily
  0712.1610}}].

\bibitem{Bouchet:2011ck}
{\scshape COrE} collaboration, \emph{{COrE (Cosmic Origins Explorer) A White
  Paper}},  \href{https://arxiv.org/abs/1102.2181}{{\ttfamily 1102.2181}}.

\bibitem{Li:2017drr}
H.~Li et~al., \emph{{Probing Primordial Gravitational Waves: Ali CMB
  Polarization Telescope}},
  \href{https://doi.org/10.1093/nsr/nwy019}{\emph{Natl. Sci. Rev.} {\bfseries
  6} (2019) 145} [\href{https://arxiv.org/abs/1710.03047}{{\ttfamily
  1710.03047}}].

\bibitem{Matsumura:2013aja}
T.~Matsumura et~al., \emph{{Mission design of LiteBIRD}},
  \href{https://doi.org/10.1007/s10909-013-0996-1}{\emph{J. Low Temp. Phys.}
  {\bfseries 176} (2014) 733}
  [\href{https://arxiv.org/abs/1311.2847}{{\ttfamily 1311.2847}}].

\bibitem{Abazajian:2019eic}
K.~Abazajian et~al., \emph{{CMB-S4 Science Case, Reference Design, and Project
  Plan}},  \href{https://arxiv.org/abs/1907.04473}{{\ttfamily 1907.04473}}.

\bibitem{Enqvist:2010vd}
K.~Enqvist, A.~Mazumdar and P.~Stephens, \emph{{Inflection point inflation
  within supersymmetry}},
  \href{https://doi.org/10.1088/1475-7516/2010/06/020}{\emph{JCAP} {\bfseries
  06} (2010) 020} [\href{https://arxiv.org/abs/1004.3724}{{\ttfamily
  1004.3724}}].

\bibitem{Hotchkiss:2011am}
S.~Hotchkiss, A.~Mazumdar and S.~Nadathur, \emph{{Inflection point inflation:
  WMAP constraints and a solution to the fine-tuning problem}},
  \href{https://doi.org/10.1088/1475-7516/2011/06/002}{\emph{JCAP} {\bfseries
  06} (2011) 002} [\href{https://arxiv.org/abs/1101.6046}{{\ttfamily
  1101.6046}}].

\bibitem{Dimopoulos:2017xox}
K.~Dimopoulos, C.~Owen and A.~Racioppi, \emph{{Loop inflection-point
  inflation}},
  \href{https://doi.org/10.1016/j.astropartphys.2018.06.002}{\emph{Astropart.
  Phys.} {\bfseries 103} (2018) 16}
  [\href{https://arxiv.org/abs/1706.09735}{{\ttfamily 1706.09735}}].

\bibitem{Okada:2016ssd}
N.~Okada and D.~Raut, \emph{{Inflection-point Higgs Inflation}},
  \href{https://doi.org/10.1103/PhysRevD.95.035035}{\emph{Phys. Rev. D}
  {\bfseries 95} (2017) 035035}
  [\href{https://arxiv.org/abs/1610.09362}{{\ttfamily 1610.09362}}].

\bibitem{Okada:2017cvy}
N.~Okada, S.~Okada and D.~Raut, \emph{{Inflection-point inflation in
  hyper-charge oriented U(1)$_X$ model}},
  \href{https://doi.org/10.1103/PhysRevD.95.055030}{\emph{Phys. Rev. D}
  {\bfseries 95} (2017) 055030}
  [\href{https://arxiv.org/abs/1702.02938}{{\ttfamily 1702.02938}}].

\bibitem{Okada:2019yne}
N.~Okada, D.~Raut and Q.~Shafi, \emph{{Inflection-Point Inflation with Axion
  Dark Matter in light of Trans-Planckian Censorship Conjecture}},
  \href{https://doi.org/10.1016/j.physletb.2020.136001}{\emph{Phys. Lett. B}
  {\bfseries 812} (2021) 136001}
  [\href{https://arxiv.org/abs/1910.14586}{{\ttfamily 1910.14586}}].

\bibitem{Bai:2020zil}
Y.~Bai and D.~Stolarski, \emph{{Dynamical Inflection Point Inflation}},
  \href{https://doi.org/10.1088/1475-7516/2021/03/091}{\emph{JCAP} {\bfseries
  03} (2021) 091} [\href{https://arxiv.org/abs/2008.09639}{{\ttfamily
  2008.09639}}].

\bibitem{Guth:2007ng}
A.H.~Guth, \emph{{Eternal inflation and its implications}},
  \href{https://doi.org/10.1088/1751-8113/40/25/S25}{\emph{J. Phys. A}
  {\bfseries 40} (2007) 6811}
  [\href{https://arxiv.org/abs/hep-th/0702178}{{\ttfamily hep-th/0702178}}].

\bibitem{Vilenkin:1983xq}
A.~Vilenkin, \emph{{The Birth of Inflationary Universes}},
  \href{https://doi.org/10.1103/PhysRevD.27.2848}{\emph{Phys. Rev. D}
  {\bfseries 27} (1983) 2848}.

\bibitem{Linde:1986fc}
A.D.~Linde, \emph{{ETERNAL CHAOTIC INFLATION}},
  \href{https://doi.org/10.1142/S0217732386000129}{\emph{Mod. Phys. Lett. A}
  {\bfseries 1} (1986) 81}.

\bibitem{Martinec:2014uva}
E.J.~Martinec and W.E.~Moore, \emph{{Modeling Quantum Gravity Effects in
  Inflation}}, \href{https://doi.org/10.1007/JHEP07(2014)053}{\emph{JHEP}
  {\bfseries 07} (2014) 053} [\href{https://arxiv.org/abs/1401.7681}{{\ttfamily
  1401.7681}}].

\bibitem{Boubekeur:2005zm}
L.~Boubekeur and D.H.~Lyth, \emph{{Hilltop inflation}},
  \href{https://doi.org/10.1088/1475-7516/2005/07/010}{\emph{JCAP} {\bfseries
  07} (2005) 010} [\href{https://arxiv.org/abs/hep-ph/0502047}{{\ttfamily
  hep-ph/0502047}}].

\bibitem{Barenboim:2016mmw}
G.~Barenboim, W.-I.~Park and W.H.~Kinney, \emph{{Eternal Hilltop Inflation}},
  \href{https://doi.org/10.1088/1475-7516/2016/05/030}{\emph{JCAP} {\bfseries
  05} (2016) 030} [\href{https://arxiv.org/abs/1601.08140}{{\ttfamily
  1601.08140}}].

\bibitem{Linde:1986fd}
A.D.~Linde, \emph{{Eternally Existing Selfreproducing Chaotic Inflationary
  Universe}}, \href{https://doi.org/10.1016/0370-2693(86)90611-8}{\emph{Phys.
  Lett. B} {\bfseries 175} (1986) 395}.

\bibitem{Susskind:2003kw}
L.~Susskind, \emph{{The Anthropic landscape of string theory}},
  \href{https://arxiv.org/abs/hep-th/0302219}{{\ttfamily hep-th/0302219}}.

\bibitem{Clifton:2007en}
T.~Clifton, A.D.~Linde and N.~Sivanandam, \emph{{Islands in the landscape}},
  \href{https://doi.org/10.1088/1126-6708/2007/02/024}{\emph{JHEP} {\bfseries
  02} (2007) 024} [\href{https://arxiv.org/abs/hep-th/0701083}{{\ttfamily
  hep-th/0701083}}].

\bibitem{Linde:2006nw}
A.D.~Linde, \emph{{Sinks in the Landscape, Boltzmann Brains, and the
  Cosmological Constant Problem}},
  \href{https://doi.org/10.1088/1475-7516/2007/01/022}{\emph{JCAP} {\bfseries
  01} (2007) 022} [\href{https://arxiv.org/abs/hep-th/0611043}{{\ttfamily
  hep-th/0611043}}].

\bibitem{Goncharov:1987ir}
A.S.~Goncharov, A.D.~Linde and V.F.~Mukhanov, \emph{{The Global Structure of
  the Inflationary Universe}},
  \href{https://doi.org/10.1142/S0217751X87000211}{\emph{Int. J. Mod. Phys. A}
  {\bfseries 2} (1987) 561}.

\bibitem{Goldwirth:1991rj}
D.S.~Goldwirth and T.~Piran, \emph{{Initial conditions for inflation}},
  \href{https://doi.org/10.1016/0370-1573(92)90073-9}{\emph{Phys. Rept.}
  {\bfseries 214} (1992) 223}.

\bibitem{Brandenberger:2016uzh}
R.~Brandenberger, \emph{{Initial conditions for inflation \textemdash{} A short
  review}}, \href{https://doi.org/10.1142/S0218271817400028}{\emph{Int. J. Mod.
  Phys. D} {\bfseries 26} (2016) 1740002}
  [\href{https://arxiv.org/abs/1601.01918}{{\ttfamily 1601.01918}}].

\bibitem{Guth:2000ka}
A.H.~Guth, \emph{{Inflation and eternal inflation}},
  \href{https://doi.org/10.1016/S0370-1573(00)00037-5}{\emph{Phys. Rept.}
  {\bfseries 333} (2000) 555}
  [\href{https://arxiv.org/abs/astro-ph/0002156}{{\ttfamily
  astro-ph/0002156}}].

\bibitem{Kobayashi:2014jga}
T.~Kobayashi and O.~Seto, \emph{{Polynomial inflation models after BICEP2}},
  \href{https://doi.org/10.1103/PhysRevD.89.103524}{\emph{Phys. Rev. D}
  {\bfseries 89} (2014) 103524}
  [\href{https://arxiv.org/abs/1403.5055}{{\ttfamily 1403.5055}}].

\bibitem{NeferSenoguz:2008nn}
V.N.~Senoguz and Q.~Shafi, \emph{{Chaotic inflation, radiative corrections and
  precision cosmology}},
  \href{https://doi.org/10.1016/j.physletb.2008.08.017}{\emph{Phys. Lett. B}
  {\bfseries 668} (2008) 6} [\href{https://arxiv.org/abs/0806.2798}{{\ttfamily
  0806.2798}}].

\bibitem{Drees:2021wgd}
M.~Drees and Y.~Xu, \emph{{Small field polynomial inflation: reheating,
  radiative stability and lower bound}},
  \href{https://doi.org/10.1088/1475-7516/2021/09/012}{\emph{JCAP} {\bfseries
  09} (2021) 012} [\href{https://arxiv.org/abs/2104.03977}{{\ttfamily
  2104.03977}}].

\bibitem{Lyth:2009zz}
D.H.~Lyth and A.R.~Liddle, \emph{{The primordial density perturbation:
  Cosmology, inflation and the origin of structure}} (2009).

\bibitem{Liddle:2003as}
A.R.~Liddle and S.M.~Leach, \emph{{How long before the end of inflation were
  observable perturbations produced?}},
  \href{https://doi.org/10.1103/PhysRevD.68.103503}{\emph{Phys. Rev. D}
  {\bfseries 68} (2003) 103503}
  [\href{https://arxiv.org/abs/astro-ph/0305263}{{\ttfamily
  astro-ph/0305263}}].

\bibitem{Tanin:2020qjw}
E.H.~Tanin and T.~Tenkanen, \emph{{Gravitational wave constraints on the
  observable inflation}},
  \href{https://doi.org/10.1088/1475-7516/2021/01/053}{\emph{JCAP} {\bfseries
  01} (2021) 053} [\href{https://arxiv.org/abs/2004.10702}{{\ttfamily
  2004.10702}}].

\bibitem{Coleman:1973jx}
S.R.~Coleman and E.J.~Weinberg, \emph{{Radiative Corrections as the Origin of
  Spontaneous Symmetry Breaking}},
  \href{https://doi.org/10.1103/PhysRevD.7.1888}{\emph{Phys. Rev. D} {\bfseries
  7} (1973) 1888}.

\bibitem{Kolb:1990vq}
E.W.~Kolb and M.S.~Turner, \emph{{The Early Universe}}, vol.~69 (1990),
  \href{https://doi.org/10.1201/9780429492860}{10.1201/9780429492860}.

\bibitem{Dufaux:2006ee}
J.F.~Dufaux, G.N.~Felder, L.~Kofman, M.~Peloso and D.~Podolsky,
  \emph{{Preheating with trilinear interactions: Tachyonic resonance}},
  \href{https://doi.org/10.1088/1475-7516/2006/07/006}{\emph{JCAP} {\bfseries
  07} (2006) 006} [\href{https://arxiv.org/abs/hep-ph/0602144}{{\ttfamily
  hep-ph/0602144}}].

\bibitem{Bernal:2021qrl}
N.~Bernal and Y.~Xu, \emph{{Polynomial inflation and dark matter}},
  \href{https://doi.org/10.1140/epjc/s10052-021-09694-5}{\emph{Eur. Phys. J. C}
  {\bfseries 81} (2021) 877}
  [\href{https://arxiv.org/abs/2106.03950}{{\ttfamily 2106.03950}}].

\bibitem{Allahverdi:2011aj}
R.~Allahverdi, A.~Ferrantelli, J.~Garcia-Bellido and A.~Mazumdar,
  \emph{{Non-perturbative production of matter and rapid thermalization after
  MSSM inflation}},
  \href{https://doi.org/10.1103/PhysRevD.83.123507}{\emph{Phys. Rev. D}
  {\bfseries 83} (2011) 123507}
  [\href{https://arxiv.org/abs/1103.2123}{{\ttfamily 1103.2123}}].

\bibitem{Starobinsky:1982ee}
A.A.~Starobinsky, \emph{{Dynamics of Phase Transition in the New Inflationary
  Universe Scenario and Generation of Perturbations}},
  \href{https://doi.org/10.1016/0370-2693(82)90541-X}{\emph{Phys. Lett. B}
  {\bfseries 117} (1982) 175}.

\bibitem{Linde:1982uu}
A.D.~Linde, \emph{{Scalar Field Fluctuations in Expanding Universe and the New
  Inflationary Universe Scenario}},
  \href{https://doi.org/10.1016/0370-2693(82)90293-3}{\emph{Phys. Lett. B}
  {\bfseries 116} (1982) 335}.

\bibitem{Munoz:2016owz}
J.B.~Mu\~noz, E.D.~Kovetz, A.~Raccanelli, M.~Kamionkowski and J.~Silk,
  \emph{{Towards a measurement of the spectral runnings}},
  \href{https://doi.org/10.1088/1475-7516/2017/05/032}{\emph{JCAP} {\bfseries
  05} (2017) 032} [\href{https://arxiv.org/abs/1611.05883}{{\ttfamily
  1611.05883}}].

\end{thebibliography}\endgroup
\end{document}